\documentclass[twocolumn,nofootinbib,showpacs,preprintnumbers,amsmath,amssymb]{revtex4}

\usepackage{graphicx}
\usepackage{dcolumn}
\usepackage{bm}
\usepackage{oldgerm}
\usepackage{psfrag}
\usepackage{color}
\usepackage{latexsym,amsfonts}
\usepackage{amsmath}
\newcommand{\be}{\begin{equation}}
\newcommand{\ee}{\end{equation}}
\newcommand{\bee}{\begin{eqnarray}}
\newcommand{\eee}{\end{eqnarray}}
\newcommand{\ds}{\displaystyle}

\begin{document}

 \title{Two-photon exchange in electron-trinucleon elastic scattering}%
\author{A.P.~Kobushkin}%
\email{kobushkin@bitp.kiev.ua}
\affiliation{%
Bogolyubov Institute for Theoretical Physics\\ Metrologicheskaya street 14B\\
03680 Kiev, Ukraine\\
}%
\affiliation{National Technical University of Ukraine ``KPI''\\
Prospect Peremogy 37, 03056 Kiev, Ukraine}
\author{Ju.V.~Timoshenko}
\email{nabla@i.ua}
\affiliation{%
Bogolyubov Institute for Theoretical Physics\\ Metrologicheskaya street 14B\\
03680 Kiev, Ukraine\\
}%
\affiliation{National Technical University of Ukraine ``KPI''\\
Prospect Peremogy 37, 03056 Kiev, Ukraine}
\date{\today}

\begin{abstract}
We discuss two-photon exchange (TPE) in elastic electron scattering off the trinucleon systems, $^3$He and $^3$H.
The calculations are done in the semirelativistic approximation with the trinucleon wave functions obtained with the Paris and CD-Bonn nucleon-nucleon potentials. An applicability area of the model is wide enough and includes the main part of kinematical domain where experimental data exist. All three TPE amplitudes (generalized form factors) for electron $^3$He elastic scattering are calculated. We find that the TPE amplitudes are few times more significant in the scattring of electrons off $^3$He then in the electron proton scattering. 
\end{abstract}

\maketitle

\section{Introduction\label{sec:Intoduction}
}
Rich information about the structure of simplest hadron systems, like the pion, proton, deuteron, tritium, $^3$He, etc., comes from measurements of electromagnetic form factors of these systems. At low $Q^2$ the form factors give information about size and form of the hadron system, while at high $Q^2$ they give information about its quark structure.

The key ingredient of such experimental studies is the extraction of the form factors from measured cross sections and polarization observables of elastic electron scattering off the hadron 
systems. Because the fine structure constant $\alpha\approx \frac1{137}$ is small one may expect that the Born approximation (the one-photon exchange) should be good enough to link the form factors with the cross sections and polarization observables.
Nevertheless, mainly due to the $G_{Ep}/G_{Mp}$ polarization measurements at Thomas Jefferson National Accelerator Facility (Jlab) \cite{Jlab.Jones,Jlab.Gayou,Punjabi}, it became clear that precise measurements of the form factors require taking into account higher order perturbative effects, such as two-photon exchange (TPE).

At present the TPE effects were analyzed in detail in the elastic $ep$ scattering, theoretically \cite{Guichon,BlundenPRL,Chen,BlundenPRC,Kondratyuk,BK_PRC_2006,BK_PRC_2007,BK_PRC_2008,BK_PRD_2009,Kivel1,Kivel2} as well as phenomenologically \cite{Arrington,Tvaskis,BorisyukKob_phen,Chen_Kao_Yang,BK_PRD_2011,Qattan}.
Besides the $ep$ scattering there are also detailed theoretical calculations of the TPE effects in the electron scattering off pions \cite{Dong_Wang,BlundenPRC_pion,BK_PRC_2011_pion} and deuterons \cite{Dong_Chen,KK-ED,KK-EDD}. 

For more complicated systems  ($^3$H, $^3$He, $^4$He) there are no systematic study of the TPE effects, with the exception of few very rough estimations \cite{BlundenPRC,Boitsov}. In Ref.~\cite{BlundenPRC} the TPE corrections to the unpolarized electron-$^3$He elastic cross-section at $Q^2$ from 1 to 6~GeV$^2$ were calculated with the elastic intermediate state only. In Ref.~\cite{Boitsov} the TPE contribution in the large angle electron-$^3$He and electron-$^4$He scattering was estimated in the multipole scattering model with simplest gaussian density in the nuclei. Lack of full THE calculations is one of the reasons why full THE effects are not taken into account in extraction procedure of the trinucleons' form factors, only part of them, the Coulomb distortion effect, is usually considered.

 
In the present paper we do the calculations of the TPE amplitude in the electron-trinucleon elastic scattering within the framework of semirelativistic approximation with the trinucleon wave function for the realistic potentials. An applicability area of our model given by Eq.~(\ref{restriction}) [see also discussion in Sec.~\ref{sec:type_II}] includes the main part of kinematical domain where experimental data exist.

The paper is organized as follows. In Sec.~\ref{sec:kinematics} we discuss kinematics and general structure of the amplitude for ultrarelativistic electron scattering off a spin $\frac12$ particle. A parametrization of the trinucleon wave function, which is used in the present calculations, is reviewed in Sec.~\ref{sec:WF2He}. The analytic expressions for the TPE amplitudes are derived in Sec.~\ref{sec:TPE_calculations}. Results of numerical calculations and conclusions are given in Sec.~\ref{sec:results}.
\section{Kinematics and general structure of the amplitude\label{sec:kinematics}}
\begin{figure}[b]
  \centering
\psfrag{k1}{$\mathbf k'$}
\psfrag{k}{$\mathbf k$}  
\psfrag{t}{$\theta$}
\psfrag{x}{$x$}    
\psfrag{z}{$z$} 
\psfrag{d}{$\mathbf P=-\mathbf q/2$}  
\psfrag{d1}{$\mathbf P\;'=\mathbf q/2$} 
  \includegraphics[height=0.2\textheight]{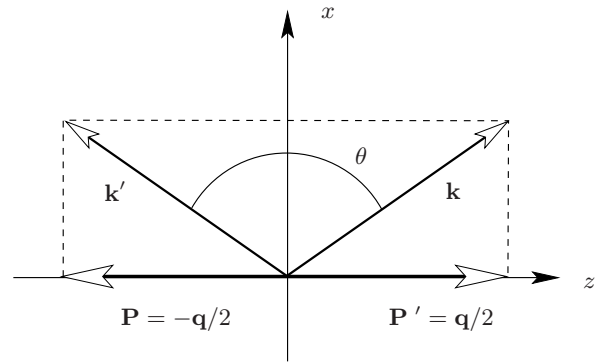}
\caption{The electron and trinucleon momenta in the Breit frame.}
\label{fig:Breit_frame}
\end{figure}

We neglect the electron mass and put the trinucleon mass $M\approx 3m$, where $m$ is the nucleon mass. The electron and 
trinucleon momenta in the initial and final states are denoted by $k$, $k'$ and $P$, $P'$; the transferred momentum 
is $q=k-k'$ and $Q^2=-q^2$. Actual calculations are done in the Breit frame (Fig.~\ref{fig:Breit_frame}), where
\begin{gather}
P_0=P'_{0}=E=\sqrt{M^2+Q^2/4}, \nonumber\\
\mathbf P_\perp=\mathbf P\,'_\perp=0, \qquad P_3=-P'_{3}=-Q/2,\nonumber\\
  q_0=q_1=q_2=0,\qquad q_3=Q,\\
k_0=k'_0=\varepsilon ,\qquad  k_1=k_1'=\varepsilon\cos\tfrac{\theta}2, \qquad k_2=k_2'=0, 
\nonumber\\
\qquad k_3=-k_3'=\tfrac12 Q=\varepsilon\sin\tfrac{\theta}2.\nonumber
\label{Breitframe}
\end{gather}
For further calculations it is useful to introduce ``plus'' and ``minus'' components of vectors according to 
$A_\pm =\sqrt{\frac12}(A_1\pm i A_2)$. 

The commonly used polarization parameter $\epsilon$ can be expressed in terms of the electron scattering angle $\theta$ in the Breit frame by
\be
\epsilon=\frac{\cos^2\tfrac{\theta}2}{1+\sin^2\tfrac{\theta}2}
\ee
(not to be confused with the components of the electron energy $\varepsilon$).

Beyond the one-photon exchange the amplitude for the electron scattering off a particle with spin $\frac12$ has the following 
structure \cite{Guichon}
\be\label{TPE}
\begin{split}
\mathcal{M}_{h\sigma'\sigma}&=\frac{4\pi\alpha}{Q^2} j^h_\mu 
\left\langle \mathbf P',\sigma'\left|H^\mu \right|\mathbf P,\sigma\right\rangle\\
&\equiv \frac{4\pi\alpha}{Q^2}j^h_\mu 
\mathcal H^{\mu}_{\sigma'\sigma},
\end{split}
\ee
where $H^\mu$ is the operator of the ``effective trinucleon current''
\be
\begin{split}
H^\mu=&\mathcal F_{1}\gamma^\mu -\mathcal F_{2}[\gamma^\mu,\gamma^\nu]\frac{q_\nu}{4M}\\
&+
\mathcal F_{3} (k+k')_\nu\gamma^\nu\frac{(P+P')^\mu}{M^2},
\label{TPE_effective}
\end{split}
\ee
with $\left|\mathbf P,\sigma\right\rangle$ and $\left|\mathbf P',\sigma'\right\rangle$ are trinucleon spinors, $\sigma$ and $\sigma'$ are $z$ projections of the trinucleon spin, $h$ is electron helicity sign, and $\mathcal F_{1}$, $\mathcal F_{2}$, and $\mathcal F_{3}$ are three independent invariant amplitudes (generalized form factors). The trinucleon spinors are normalized by
\be\label{normalization}
\left\langle \mathbf P,\sigma|\mathbf P,\sigma\right\rangle=2M.
\ee
In the Breit frame components of the electromagnetic current of the electron $j^h_\mu=\bar u_h(k')\gamma_\mu u_h(k)$ 
[with $u_h(k)$ and $u_h(k')$ being spinors of the initial and final electrons, respectively] are the following 
\be\label{electron_em_current}
\begin{split}
 j^h_0=2\varepsilon\cos\tfrac{\theta}2, \quad j^h_\pm=\sqrt2 \varepsilon(1\mp h\sin\tfrac{\theta}2),\quad
 j^h_3=0.
\end{split}
\ee

In general case all generalized form factors $\mathcal F_{1}$, $\mathcal F_{2}$, and $\mathcal F_{3}$ are complex functions of two independent
variables, e.g. $Q^2$ and $\epsilon$. At the zeroth order in $\alpha$ the form factors $\mathcal F_{1}$ and $\mathcal F_{2}$ reduce to the Dirac and Pauli form factors $F_1(Q^2)$ and $F_2(Q^2)$, while the form factor $\mathcal F_{3}$ vanishes. At the first order in $\alpha$ all of them are nonvanishing.

Instead of the Dirac and Pauli form factors $F_1(Q^2)$ and $F_2(Q^2)$, one usually introduces their linear combinations (the charge and magnetic form factors, respectively)
\be\label{charge_and_magnetic}
\begin{split}
 & G_C(Q^2)=Z^{-1}\left[F_1(Q^2) - \eta F_2(Q^2)\right],\\
 & G_M(Q^2)=\frac{m}{\mu M}\left[F_1(Q^2) + F_2(Q^2)\right],
\end{split}
\ee
which ``diagonalize'' the unpolarized cross section
$$
\frac{d\sigma}{d\Omega} \sim \epsilon G_C^2(Q^2) + Z^{-2}\left(\dfrac{\mu M}{m}\right)^2 \eta G_M^2(Q^2)
$$
and satisfy $G_C(0)=G_M(0)=1$. In Eq.~(\ref{charge_and_magnetic}) $\eta=Q^2/(4M^2)$, $Z$ is the trinucleon charge in units of the elementary charge $e$, and  $\mu$ is the trinucleon magnetic moment in nuclear magnetons, $\mu(^3\mathrm{He})\approx -2.127$ and $\mu(^3\mathrm{H})\approx 2.978$. Beyond the one-photon exchange we can also introduce (see later) appropriate linear combination of the generalized form factors, which diagonalizes the cross section up to second order perturbative corrections \cite{BK_PRC_2007}.

In the Breit frame matrix elements of the effective trinucleon current (\ref{TPE_effective}) are
\be\label{components_of_effectiv}
\begin{split}
 \mathcal H_{\sigma'\sigma 0}
=&
  2M\left( \mathcal F_1 - \eta \mathcal F_2 +\frac{E \varepsilon}{M^2}\mathcal F_3\right)\delta_{\sigma'\sigma}
   \\
& \pm \cos\frac{\theta}2\frac{E \varepsilon Q}{M^2}\mathcal F_3
\delta_{\sigma',\sigma\pm1},\\
\mathcal H_{\sigma'\sigma\pm}
=& \pm \sqrt2 Q\left( \mathcal F_1 +\mathcal F_2\right) 
\delta_{\sigma',\sigma\pm1},\\
 \mathcal H_{\sigma'\sigma 3} 
=&0.
\end{split}
\ee

It is useful to introduce the reduced amplitude $T_{h\sigma'\sigma}$ instead of the usual amplitude $\mathcal{M}_{h\sigma'\sigma}$:
\be\label{reduce_ampl}
\mathcal{M}_{h\sigma'\sigma}=\frac{16\pi\alpha}{Q^2}\varepsilon M T_{h\sigma'\sigma}.
\ee

From Eqs.~(\ref{TPE}), (\ref{electron_em_current}), and (\ref{components_of_effectiv}) we get explicitly the spin structure 
of the amplitude 
\be\label{spin_structure}
 T_h=\left(
\begin{array}{ll}
Z \cos\frac{\theta}2 \mathcal G_C          & \sqrt\eta(f_1 - h \sin\frac{\theta}2 f_2)\\[0.25cm]
-\sqrt\eta(f_1 + h \sin\frac{\theta}2 f_2) & Z \cos\frac{\theta}2 \mathcal G_C
\end{array}
\right),
\ee
where
\be\label{where}
\begin{split}
 & \mathcal G_C=Z^{-1}\left(\mathcal F_1 - \eta \mathcal F_2 + \frac{\varepsilon E}{M^2}\mathcal F_3\right),\\
 & f_1=\mathcal F_1 + \mathcal F_2 +\frac{\varepsilon E}{M^2}\mathcal F_3\cos^2\frac{\theta}2,\\
 & f_2=\mathcal F_1 + \mathcal F_2.
\end{split}
\ee

The differential cross section is given by 
\be\label{dif_cs}
\frac{d\sigma}{d\Omega}=\frac{d\sigma_\mathrm{Mott}}{d\Omega}\frac{\sigma_0}{\epsilon(1+\eta)},
\ee
where 
\be\label{reduced_cs}
\begin{split}
 \sigma_0=&\epsilon |\mathcal G_C(Q^2,\epsilon)|^2 \\
 &+ 
 Z^{-2}\left(\dfrac{\mu M}{m}\right)^2 \eta\left( 1 + 2\tan^2\frac{\theta}2\right)|\mathcal G_M(Q^2,\epsilon)|^2 \\
 &+
 Z^{-2}\eta\epsilon^2\frac{1-\epsilon}{1+\epsilon}|\mathcal G_3(Q^2,\epsilon)|^2
 \end{split}
\ee
is so called reduced cross section and
\be\label{Mott}
\begin{split}
&\frac{d\sigma_\mathrm{Mott}}{d\Omega}
=\left(\frac{\alpha Z \cos\tfrac12\theta_\mathrm{lab}}{2E_\mathrm{lab}\sin^2\tfrac12\theta_\mathrm{lab}}\right)^2 \\
 & \times \frac{1}{1+(2E_\mathrm{lab}/M)\sin^2\tfrac12\theta_\mathrm{lab}},
\end{split}
\ee
is the Mott cross section. In Eqs.~(\ref{reduced_cs}) and (\ref{Mott}) $E_\mathrm{lab}$ and $\theta_\mathrm{lab}$ are the electron energy and scattering angle in the laboratory frame, $\mathcal G_C$ is given by the first line in Eq.~(\ref{where}), and
\be\label{Magnetic_generalized_and_G3}
\begin{split}
&\mathcal G_M = \frac{m}{\mu M}\left(\mathcal F_1 + \mathcal F_2 + \frac{\epsilon \varepsilon E}{M^2}\mathcal F_3\right),\\
&\mathcal G_3 =\frac{\varepsilon E}{4M^2}\mathcal F_3 .
\end{split}
\ee

Performing the $\alpha$-expansion of the reduced cross section we obtain
\be\label{reduced_cs_alpha-expansion}
\begin{split}
 \sigma_0 & = \epsilon |\mathcal G_C(Q^2,\epsilon)|^2 \\
 &+ 
 Z^{-2}\left(\dfrac{\mu M}{m}\right)^2 \eta\left( 1 + 2\tan^2\frac{\theta}2\right)|\mathcal G_M(Q^2,\epsilon)|^2 \\
 &+\mathcal O(\alpha^2)
 \end{split}
\ee
where 
$|\mathcal G_C|^2 \cong G_C^2 +2 G_C\mathrm{Re}\Delta\mathcal G_E$, and 
$|\mathcal G_M|^2 \cong G_M^2 +2 G_M\mathrm{Re}\Delta\mathcal G_M$ with $\Delta\mathcal G_C$ and $\Delta\mathcal G_M$ being the TPE corrections of order $\alpha$ to the corresponding form factors. The obtained expression for the cross section is similar to the Rosenbluth cross section, where $\mathcal G_C(Q^2,\epsilon)$ and $\mathcal G_M(Q^2,\epsilon)$ play role of the charge and magnetic form factors. Thus later they are called generalized charge and magnetic form factors (see Ref.~\cite{BorisyukKob_phen}).

\section{Trinucleon wave function\label{sec:WF2He}}
We use the parametrization~\cite{Baru} of the totally antisymmetric trinucleon wave function calculated with the Paris~\cite{Paris} and cd-Bonn~\cite{cd-Bonn} potentials. This parametrization is restricted to five partial waves,
\be
\label{Channelspin}
\left|\left[\left((\ell s)j\tfrac12\right)KL\right]\tfrac12\right>,
\ee
where $\ell$, $j$, and  $s$ are the orbital, total, and spin angular momenta
for the pair (the $2^{\text{nd}}$ and $3^{\text{rd}}$ nucleons), and $L$ and $K$ are
the relative orbital angular momentum for the spectator (the $1^{\text{st}}$
nucleon) and the so called channel spin, respectively. The appropriate quantum numbers of the partial waves are collected
in Table~I of Ref.~\cite{Baru}.

The standard definition of the Jacobi coordinates for the three-particle system $\mathbf r$ and $\boldsymbol \rho$ and the corresponding momenta $\boldsymbol\mu$ and  $\boldsymbol \nu$ is 
\bee
\label{internal_mom}
\begin{array}{ll}
 \mathbf r_1=\mathbf R + \tfrac23\boldsymbol\rho, & \mathbf p_1=\tfrac13 \mathbf P +\boldsymbol\nu,  \\[0.25 cm]
 \mathbf r_2=\mathbf R - \tfrac13\boldsymbol\rho + \tfrac12\mathbf r, & \mathbf p_2=\tfrac13 \mathbf P -\tfrac12\boldsymbol\nu +  \boldsymbol \mu, \\[0.25 cm]
 \mathbf r_3=\mathbf R - \tfrac13\boldsymbol\rho - \tfrac12\mathbf r, & \mathbf p_3=\tfrac13 \mathbf P -\tfrac12\boldsymbol\nu -  \boldsymbol \mu.
\end{array}
\eee
Here $\mathbf R$ is the coordinate of the trinucleon center of mass and $\mathbf P$ is the trinucleon momentum.

Explicitly, the trinucleon wave function is (see Ref.~\cite{KobStrok})
\begin{widetext}
\bee \label{WF_MOMENTUM}
\Psi_{\sigma T_3}(\boldsymbol\mu, \boldsymbol \nu\,)=
&&\sum_\xi\left\{\frac{1}{4\pi}\delta_{\xi\sigma}
\sum_{\tau_3,t_3}
\left<1\tfrac12 \tau_3t_3\bigm|\tfrac12 T_3\right>\psi_1(\mu,\nu)
\left|00;1\tau_3\right>\chi_{\xi t_3}+
\sum_{s_3}
\left[\frac{1}{4\pi}\left<1\tfrac12s_3\xi\bigm|\tfrac12\sigma\right> \psi_2(\mu,\nu)\right.\right.\nonumber\\
&&\left.\left. -\sqrt{\frac{1}{4\pi}}\sum_{L_3K_3}\left<1\tfrac12s_3\xi\bigm|\tfrac32 K_3\right>
\left<\tfrac32 2 K_3 L_3 \bigm|\tfrac12\sigma\right> Y_{2L_3}(\widehat {\boldsymbol \nu}\,)\psi_3(\mu,\nu) \right.\right.\nonumber\\
&&\left.\left. -\sqrt{\frac{1}{4\pi}}\sum_{\ell_3M}\left<12 s_3\ell_3|1M\right>
\left<1\tfrac12M\xi\bigm|\tfrac12 \sigma\right>  Y_{2\ell_3}(\widehat {\boldsymbol \mu}\,)\psi_4(\mu,\nu)\right.\right.
\label{Explicity}\\
&&\left.\left.
+\sum_{\ell_3ML_3K_3}\left<12 s_3\ell_3|1M\right>
\left<1\tfrac12M\xi\bigm|\tfrac32 K_3\right>
\left<\tfrac32 2 K_3 L_3\bigm|\tfrac12\sigma \right> 
 Y_{2\ell_3}(\widehat {\boldsymbol \mu}\,)Y_{2L_3}(\widehat {\boldsymbol \nu}\,)\psi_5( \mu, \nu)
\right] \left|1s_3;00\right>\chi_{\xi T_3}\right\}\  ,\nonumber
\eee
where $\sigma$ and $\xi$ are the spin projections of the trinucleon and the spectator nucleon, $T_3$, $t_3$, and $\tau_3$ are
the isospin projections of the trinucleon, the spectator nucleon, and the pair; $M$ is the projection of the total angular momentum of the pair; $\chi_{\xi t_3}$ and $\left|ss_3;\tau\tau_3\right>$ are the spin-isospin wave function of the spectator nucleon and the pair; $s_3$, $L_3$, $K_3$, and $\ell_3$ are $z$~projections of appropriate momenta; $\left< j_1j_2m_1m_2\bigm|JJ_z\right>$ are the Clebsch-Gordan coefficients.

Later on we  will also need the trinucleon wave function in the coordinate space
\bee \label{WF_COORDINATE}
\Phi_{\sigma T_3}(\mathbf r, \boldsymbol \rho\,)=
&&(r\rho)^{-1}\sum_\xi\left\{\frac{1}{4\pi}\delta_{\xi\sigma}
\sum_{\tau_3,t_3}
\left<1\tfrac12 \tau_3t_3\bigm|\tfrac12 T_3\right>\phi_1(r,\rho)
\left|00;1\tau_3\right>\chi_{\xi t_3}+
\sum_{s_3}
\left[\frac{1}{4\pi}\left<1\tfrac12s_3\xi\bigm|\tfrac12\sigma\right> \phi_2(r,\rho)\right.\right. \nonumber\\
&&\left.\left. +\sqrt{\frac{1}{4\pi}}\sum_{L_3K_3}\left<1\tfrac12s_3\xi\bigm|\tfrac32 K_3\right>
\left<\tfrac32 2 K_3 L_3 \bigm|\tfrac12\sigma\right> Y_{2L_3}(\widehat {\boldsymbol \rho}\,)\phi_3(r,\rho) \right.\right. \nonumber\\
&&\left.\left. +\sqrt{\frac{1}{4\pi}}\sum_{\ell_3M}\left<12 s_3\ell_3|1M\right>
\left<1\tfrac12M\xi\bigm|\tfrac12 \sigma\right>  Y_{2\ell_3}(\widehat {\mathbf r}\,)\phi_4(r,\rho)\right.\right. \label{Explicity_r}\\
&&\left.\left.
+\sum_{\ell_3ML_3K_3}\left<12 s_3\ell_3|1M\right>
\left<1\tfrac12M\xi\bigm|\tfrac32 K_3\right>
\left<\tfrac32 2 K_3 L_3\bigm|\tfrac12\sigma \right> 
 Y_{2\ell_3}(\widehat {\mathbf r}\,)Y_{2L_3}(\widehat {\boldsymbol \rho}\,)\phi_5( r, \rho)
\right] \left|1s_3;00\right>\chi_{\xi T_3}\right\},  \nonumber
\eee
\end{widetext}
where $\phi_n(r,\rho)$ are partial radial wave functions given by
\be\label{Fourier_transform}
\phi_n(r,\rho)=\frac2{\pi}\int d\mu d\nu \mu^2 \nu^2 j_l(\mu r) j_L(\nu \rho)\psi_n(\mu,\nu).
\ee
Here $j_\ell(z)$ are spherical Bessel functions (not to be confused with the components of the electron current $j_\mu^h$).
\begin{figure}[t]
  \centering
\psfrag{e}{$e$} 
\psfrag{e'}{$e'$}  
\psfrag{1}{$1$}
\psfrag{2}{$2$}    
\psfrag{3}{$3$}   
\psfrag{a}{(a)}   
\psfrag{b}{(b)}   
\psfrag{c}{(c)}
  \includegraphics[height=0.150\textheight]{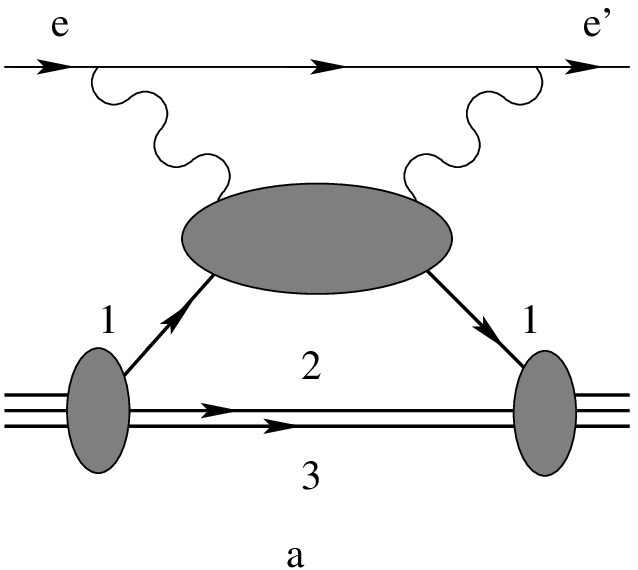}\\[0.25cm]
  \includegraphics[height=0.125\textheight]{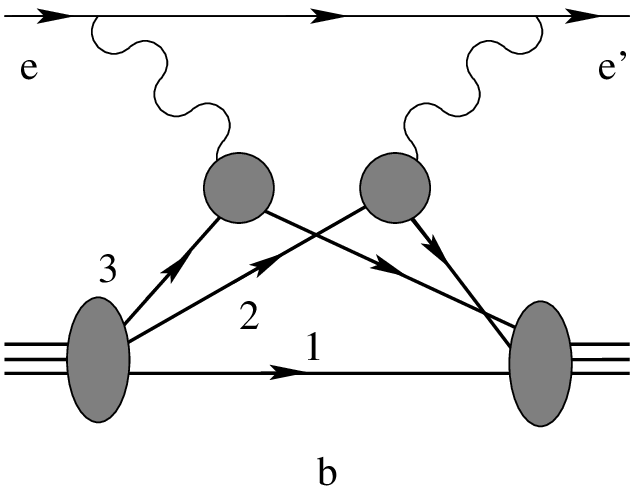}
  \includegraphics[height=0.125\textheight]{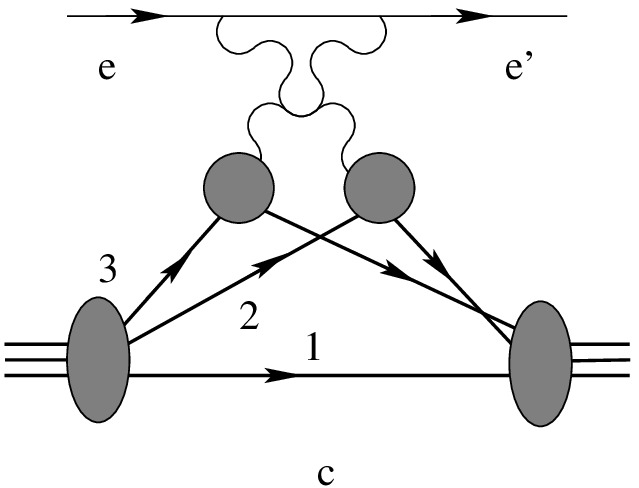}
\caption{Diagrams of the type I (a) and the type II (b and c).}
\label{fig:Diagrams}
\end{figure}
\section{Calculation of the TPE amplitude\label{sec:TPE_calculations}}
Similarly to the $ed$~scattering \cite{KK-ED} we consider only TPE diagrams
where virtual photons interact directly with the nucleons (in our model we exclude nonnucleon degrees of freedom, such as mesons, isobars, etc.). There are two types of such diagrams and the total amplitude is given by
$$\Delta \mathcal M=\Delta \mathcal M^\mathrm{I
} + \Delta \mathcal M^\mathrm{II}.$$  
The amplitude $\Delta \mathcal M^\mathrm{I}$ corresponds to the diagram  Fig.~\ref{fig:Diagrams}(a) where both photons interact with the same nucleon. The amplitude of the second type, $\Delta \mathcal M^\mathrm{II}$, corresponds to the sum of diagrams Figs.~\ref{fig:Diagrams}(b) and (c) where the photons interact with different nucleons. 
 
Below we will consider the TPE corrections to the generalized form factors of $^3$He. The corrections to the generalized form factors of $^3$H are obtained from them by interchange $p \leftrightarrow n$.
\subsection{Diagrams of type I}
The amplitude for the electron-nucleon scattering has a structure similar to (\ref{TPE})
\be\label{TPE_N}
\begin{split}
\mathcal{M}_{2\gamma}^N&=\frac{4\pi\alpha}{Q^2}j^h_\mu
\left\langle \mathbf p{\,'}\xi'\left|H_N^\mu \right|\mathbf p\,\xi\right\rangle ,
\end{split}
\ee
where $\mathbf p$, $\mathbf p'$ and $\xi$, $\xi'$ are nucleon momenta and spin projections in the initial and final 
states, respectively, and  $H^\mu_N$ is the operator of the ``effective current'' of the nucleon
\be
\begin{split}
H^\mu_N=&\mathcal F_{1N}\gamma^\mu -\mathcal F_{2N}[\gamma^\mu,\gamma_\nu]\frac{q^\nu}{4m}\\
&+\mathcal F_{3N} K_\nu\gamma^\nu\frac{(p+p')^\mu}{M^2}.
\end{split}
\label{TPE_effective1}
\ee
$\mathcal F_{1N}$, $\mathcal F_{2N}$, and $\mathcal F_{3N}$ are the generalized form factors of the elastic $eN$~scattering. 

Using these formulae we easily get the TPE correction of the type I to the amplitude:
\be\label{eff_I}
\begin{split}
 \Delta T_{h\sigma'\sigma}^\mathrm I=&\frac{3}{4kM} j^{h\mu}
 \Delta \mathcal H^{\mathrm Ia}_{\sigma'\sigma\mu},\\
 \Delta \mathcal H^{\mathrm I}_{\sigma'\sigma\mu}=&
 6M\int d^3\mu d^3\nu \Psi^\dag_{\sigma'}(\boldsymbol\mu, \boldsymbol \nu + \tfrac23 \mathbf q\,)\\
 & \times \Delta H_{N_1\mu}\Psi_{\sigma}(\boldsymbol\mu, \boldsymbol \nu\,), 
\end{split}
\ee
where $N_1$ means the spectator nucleon and $\Delta H_{N_1\mu}$ is the part of its effective current (\ref{TPE_effective1}) proportional to $\alpha$. The factor of 3 in Eq.~(\ref{eff_I}) is spectroscopical factor squared. 

As a result the corrections of the type I to the $^3$He generalized form factors are
\be\label{FFs_I}
\begin{split}
  \Delta \mathcal G_E^{\mathrm I} = & Z^{-1}\left\{\left(\Delta \mathcal G_{Ep} + 
  2\Delta \mathcal G_{En}\right)I_{11}^0\left(Q^2\right) \right.\\
  & \left.+
  3\Delta \mathcal G_{Ep}\left[
  I_{22}^0\left(Q^2\right) + I_{33}^0\left(Q^2\right) \right.\right.\\ 
  & \left.\left. + I_{44}^0\left(Q^2\right) + I_{55}^0\left(Q^2\right)
  \right]\right\},\\
  \Delta \mathcal G_E^{\mathrm I} = & \mu^{-1}\left( \Delta\mathcal F^\mathrm{I}_1 + \Delta\mathcal F^\mathrm{I}_2 +
  \frac{\epsilon kE}{M^2}\mathcal F^\mathrm{I}_3\right),
\end{split}
\ee
where
\be\label{FFs_I.a}
\begin{split}
  \Delta \mathcal F_1^{\mathrm I} +\Delta \mathcal F_2^{\mathrm I} = & 
  \left(\Delta \mathcal G_{Mp} + 2\Delta \mathcal G_{Mn}\right)I_{11}^0\left(Q^2\right) \\
  & -
  \Delta \mathcal G_{Mp}\left[
  I_{22}^0\left(Q^2\right) + I_{33}^0\left(Q^2\right) \right.\\
  & \left. + I_{44}^0\left(Q^2\right) + I_{55}^0\left(Q^2\right)  \right.\\
  & \left.+2\sqrt{2}I_{23}^2\left(Q^2\right) +2\sqrt{2} I_{45}^2\left(Q^2\right)\right.\\
  & \left. - I_{55}^2\left(Q^2\right)
  - I_{33}^2\left(Q^2\right)
  \right],\\
  \mathcal F_3^{\mathrm I} = & \frac{M^2}{kE}\left\{\left(\mathcal F_{3 p} + 2\mathcal F_{3 n}\right)I_{11}^0\left(Q^2\right) \right.\\
  & \left. -
  \mathcal F_{3 p}\left[
  I_{22}^0\left(Q^2\right) + I_{33}^0\left(Q^2\right)  \right.\right.\\
  & \left.\left.+ I_{44}^0\left(Q^2\right) + I_{55}^0\left(Q^2\right)  
  \right. \right.\\
  & \left.\left.+2\sqrt{2}I_{23}^2\left(Q^2\right) + 2\sqrt{2} I_{45}^2\left(Q^2\right)  \right.\right.\\
  & \left.\left.- 
  I_{55}^2\left(Q^2\right) - I_{33}^2\left(Q^2\right)
  \right]\right\},\\
\end{split}
\ee
and
\be\label{body_ff1}
I_{nm}^\ell(Q^2)=\int dr d \rho \,j_\ell\left(\tfrac23 Q\rho\right)\phi_n(r,\rho)\phi_m(r,\rho),
\ee
where $\phi_n(r,\rho)$ are partial radial wave functions of the trinucleon in the coordinate space, see Eq.~(\ref{Fourier_transform}).
\subsection{Diagrams of type II \label{sec:type_II}}
Let us introduce four dimensional ``internal'' momenta in the trinucleon in the initial and final states
\be\label{II.4D_internal_momenta}
\begin{array}{ll}
\nu=\tfrac23\left[p_1-\tfrac12\left(p_2 + p_3\right)\right],& 
\nu'=\tfrac23\left[p_1'-\tfrac12\left(p_2' + p_3'\right)\right],\\[0.25cm]
\mu=\tfrac12\left(p_2 - p_3\right),& \mu'=\tfrac12\left(p_2' - p_3'\right),
\end{array}
\ee
where $p_1$, $p_2$, $p_3$ and $p_1'$, $p_2'$, $p_3'$ are the nucleon four momenta. For the diagrams Fig.~\ref{fig:Diagrams}(b) and (c)  
 $p_1=p_1'$ and $\nu$ and $\nu'$ are connected by
\be\label{mu}
\nu'=\nu-\tfrac13 q.
\ee

The corresponding amplitude is given by the sum of the two diagrams 
$\Delta\mathcal{M}^{\mathrm{II}}=\Delta\mathcal{M}^{\mathrm{II}}_{\mathrm{b}}+\Delta\mathcal{M}^{\mathrm{II}}_{\mathrm{c}}$. Thus
\begin{equation}
\begin{split}
 i\Delta\mathcal{M}^{\mathrm{II}}_{h\sigma'\sigma}=&3 i \int \frac{d^4\nu}{(2\pi)^4}\frac{d^4\mu}{(2\pi)^4}\frac{d^4\mu\,'}{(2\pi)^4}\\
 & \times t_{\mu\nu}^h
 G(\Delta_1,\Delta_2)\frac{{\mathcal{T}}_{\sigma'\sigma}^{\mu\nu}}{D}\\
 \equiv & 3i\int t^h_{\mu\nu}\,dT
 _{\sigma'\sigma}^{\mu\nu} \;.
\end{split}
\label{TPE.2}
\end{equation}
Here the factor of 3 is spectroscopical factor squared, $\Delta_1=k-l$ and $\Delta_2=l-k'$ are the four momenta of the photons ($l$ is the momentum of the intermediate electron) and 
\bee
t^h_{\mu\nu}=&&(-ie)^2
 \frac{\tau_{\mu\nu}^h}{l^2+i0}\;  ,
\label{Nucleon_propagator1}\\
 G(\Delta_1,\Delta_2)=&&
 \displaystyle{\frac{ -i}{\Delta_1^2+i0}\cdot\frac{ -i}{\Delta_2^2+i0}},
\label{Nucleon_propagator2} \\
 \mathcal{T}^{\mu\nu}_{\sigma'\sigma}
 =&&
 (ie)^2 
 \left< \mathbf P',\sigma'\right|
 iV^\dag(p_1,p'_2,p'_3) 
\nonumber\\
 && \times i(p\!\!\!/_1+m)i(p\!\!\!/'_2+m)i(p\!\!\!/'_3+m) \label{Nucleon_propagator3} \\
 && \times
 \Gamma^\mu_{N_2}\left(\Delta_1\right)\Gamma^\nu_{N_3}\left(\Delta_2\right) i(p\!\!\!/_2+m) \nonumber\\
 &&\times i(p\!\!\!/_3+m)iV(p_1,p_2,p_3)
 \left|\mathbf P,\sigma\right>
 ,\nonumber\\
 D=&&(p_1^2-m^2+i0)\nonumber\\
 &&\times({p'}_2^2-m^2+i0)(p_2^2-m^2+i0)\label{Nucleon_propagator4} \\
 &&\times(p_3^2-m^2+i0)({p'}_3^2-m^2+i0),\nonumber
\eee
where
\be\label{lepton_line}
\begin{split}
\tau_{\mu\nu}^h=&\bar u_h(k')\left[ \gamma_\mu l\!\!/ \gamma_\nu + \gamma_\nu l\!\!/ \gamma_\mu\right] u_h(k)  
\\
=&j^h_\nu(k+k')_\mu +  j^h_\mu(k+k')_\nu \\
& +
\tfrac12\bar u_h(k')\left[\gamma_\mu(-\Delta\!\!\!\!/_1 + \Delta\!\!\!\!/_2)\gamma_\nu \right.\\
&  \left. +\gamma_\nu(-\Delta\!\!\!\!/_1 + \Delta\!\!\!\!/_2)\gamma_\mu \right] u_h(k)
\end{split}
\ee 
and 
$V(p_1,p_2,p_3)$ and $V(p_1,p'_2,p'_3)$ are the trinucleon vertex functions, which are connected with the trinucleon wave function by Eq.~(\ref{IA.4});
\be\label{gamma_NN}
\Gamma^\mu_{N_i}\left(\Delta\right)=\gamma^\mu_i F_{1N_i}(-\Delta^2) - 
\frac{\Delta_\nu}{4m}[\gamma^\mu_i,\gamma^\nu_i] F_{2N_i}(-\Delta^2)
\ee
is the electromagnetic current for $i$-th nucleon [$F_{1N_i}(-\Delta^2)$ and $F_{2N_i}(-\Delta^2)$ are the Dirac and Pauli 
form factors of the nucleon $N_i$], $p\!\!\!/_i\equiv p_i^\rho \gamma_{i\rho}$, and $\gamma_{i\rho}$ are the Dirac matrices for the $i$-th nucleon.

\begin{figure}[t,b]
  \centering
  \psfrag{a}{(a)}
  \psfrag{b}{(b)}
  \psfrag{c}{(c)}
  \psfrag{d}{(d)}
  \psfrag{1}{1}\psfrag{2}{2}\psfrag{3}{3}\psfrag{2'}{$3'$}\psfrag{3'}{$2'$}
  \includegraphics[height=0.225\textheight]{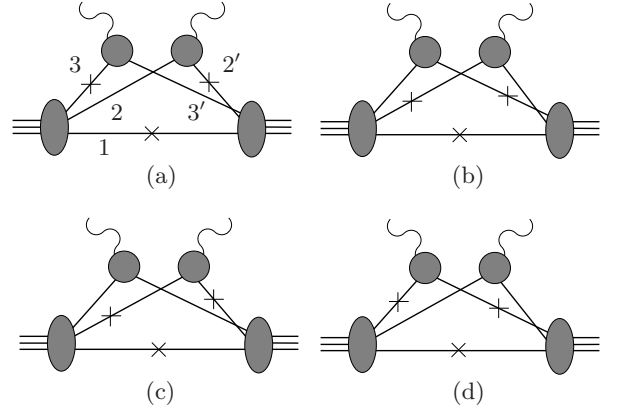}
\caption{The poles taken into account in the integration over $d\mu_0d\nu_0d\nu'_0$.}
\label{fig:Poles}
\end{figure}

Integrating over $d\nu_0d\mu_0d\mu'_0$ one has to take into account four types of poles shown in Fig.~\ref{fig:Poles} and the amplitude $\Delta\mathcal{M}^{\mathrm{II}}_{h\sigma'\sigma}$ becomes a sum of four respective terms
\be\label{a-b-c-d}
\begin{split}
 \Delta\mathcal{M}^{\mathrm{II}}_{h\sigma'\sigma}& =
 \frac{16\pi \alpha}{Q^2}\varepsilon M\left(
 \Delta T^{\mathrm{II}\,(a)}_{h\sigma'\sigma} + \Delta T^{\mathrm{II\, (b)}}_{h\sigma'\sigma} 
  + \Delta T^{\mathrm{II\, (c)}}_{h\sigma'\sigma} + \Delta T^{\mathrm{II\, (d)}}_{h\sigma'\sigma}
 \right)\\
 & =
3 \int t^h_{\mu\nu}(dT^{\mathrm{(a)}\mu\nu}_{\sigma'\sigma} + dT^{\mathrm{(b)}\mu\nu}_{\sigma'\sigma} 
+ dT^{\mathrm{(c)}\mu\nu}_{\sigma'\sigma} + dT^{\mathrm{(d)}\mu\nu}_{\sigma'\sigma}).
\end{split}
\ee
What follows is a discussion of the contribution coming from the poles of the diagram Fig.~\ref{fig:Poles}~(a). Other contributions are calculated similarly.
\be
\begin{split}\label{Pole_int}
dT^{\mathrm{(a)}\mu\nu}_{\sigma'\sigma} 
 \to & -i\frac{d^3\nu d^3\mu d^3\mu'}{2E_1 2E_2 2E'_3 (2\pi)^9} \\
& \times\frac{\mathcal{T}^{\mathrm{(a)}\mu\nu}_{\sigma'\sigma}}{({p_2'}^2-m^2+i0 )(p_3^2-m^2+i0)}.
\end{split}
\ee
Here $\mathcal{T}^{\mathrm{(a)}\mu\nu}_{\sigma'\sigma}$ are given in (\ref{Nucleon_propagator3})
where three nucleons marked by a cross in Fig.~\ref{fig:Poles}~(a) are on mass shell;
\begin{equation}\label{Pole_integration}
\begin{split}
E_1=&E_1'=\sqrt{m^2+\left(\boldsymbol\nu-\tfrac16\mathbf q\;\right)^2},\\ 
E'_2=&\sqrt{m^2+\left(\tfrac12\boldsymbol\nu' - \boldsymbol\mu' - \tfrac16\mathbf q\;\right)^2},\\
E_3=&\sqrt{m^2+\left(\tfrac12\boldsymbol\nu + \boldsymbol\mu +\tfrac16\mathbf q\;\right)^2},
\end{split}
\end{equation}
$E'_2$ and $E_3$ are obtained from energy conservation in the trinucleon vertexes, and $\boldsymbol \nu'=\boldsymbol \nu -\frac13 \mathbf q$. 

Now we can introduce the wave function of the trinucleon,
which has two nucleons on the mass shell [see Eq.~(\ref{IA.4})], and $dT^{\mathrm{(a)}\mu\nu}_{\sigma'\sigma}$ becomes
\bee\label{dT}
&&dT_{\sigma'\sigma}^{\mathrm{(a)}\mu\nu}= \frac{e^2 M}{2m^2}\frac{d^3\nu d^3\mu d^3\mu'}{(2\pi)^32E_12E_32E'_2} \\
&&\times\hspace{-0.25cm} \sum_{\xi_1,\xi_3,\xi'_2}\hspace{-0.25cm}
\Psi^\dag_{\sigma'} (\boldsymbol \mu',\boldsymbol \nu')
|\mathbf p_1,\xi_1\rangle |\mathbf p'_2,\xi'_2\rangle\langle\mathbf p'_2,\xi'_2|
\Gamma_{N_2\nu }^\nu (p\!\!\!/_2+m) \nonumber \\
&& \times (p\!\!\!/_3'+m)\Gamma_{N_3}^\mu
|\mathbf p_3,\xi_3\rangle\langle\mathbf p_3,\xi_3|\langle\mathbf p_1,\xi_1|
\Psi_{\sigma} (\boldsymbol \mu,\boldsymbol \nu), \nonumber
\eee
where $|\mathbf p_1,\xi_1\rangle$, $|\mathbf p_3,\xi_3\rangle$ and $|\mathbf p'_2,\xi'_2\rangle$ are the Dirac spinors of corresponding nucleons.
 
The space components of the nucleon momenta are given by 
\be\label{nucleon_momenta}
\begin{array}{ll}
\mathbf p_1  =  \mathbf p'_1 \equiv \boldsymbol \upsilon=\boldsymbol \nu - \tfrac16 \mathbf q,\\[0.25cm]
\mathbf p_2  = -\tfrac14 \mathbf q - \tfrac12 \boldsymbol \upsilon + \boldsymbol \mu,&
\mathbf p'_2 =  \tfrac14 \mathbf q - \tfrac12 \boldsymbol \upsilon + \boldsymbol \mu',\\[0.25cm]
\mathbf p_3  = -\tfrac14 \mathbf q - \tfrac12 \boldsymbol \upsilon - \boldsymbol \mu,&
\mathbf p'_3 =  \tfrac14 \mathbf q - \tfrac12 \boldsymbol \upsilon - \boldsymbol \mu'.
\end{array}
\ee
Later on we will assume that 
\be\label{approximation}
|\boldsymbol \mu|,\; |\boldsymbol \mu'|,\,\; |\boldsymbol \upsilon|\ll m. 
\ee
We will also restrict our calculations to 
\be\label{condition}
Q \ll 4 m
\ee

Keeping only terms linear in $\boldsymbol \mu$, $\boldsymbol \mu'$, and $\boldsymbol \upsilon$ we get
\begin{equation}\label{Pole_int_approx}
\begin{split}
E_1=&E_1' = \sqrt{m^2+\boldsymbol\upsilon^2}\approx m,\\ 
E_3 \approx &m +\frac{Q^2}{32m} +\frac{Q(\upsilon_3 + 2 \mu_3)}{8m},
\\
E_2= & E-E_1-E_3\approx m+\frac{Q^2}{96 m} -\frac{Q(\upsilon_3 + 2\mu_3)}{8m} ,\\
E'_2 \approx & m+\frac{Q^2}{32m} - \frac{Q(\upsilon_3-2\mu'_3)}{8m} ,
 \\
 E'_3 = & E-E'_1-E'_2 \approx m+\frac{Q^2}{96 m} + \frac{Q (\upsilon_3 - 2\mu'_3)}{8m}.
\end{split}
\end{equation}
and 
\be\label{photon_momenta}
\begin{split}
 \Delta_{1}=&\left( \Delta_0, \tfrac12 \mathbf q +\boldsymbol \mu - \boldsymbol \mu'\right),\\
 \Delta_{2}=&\left(-\Delta_0, \tfrac12 \mathbf q -\boldsymbol \mu + \boldsymbol \mu'\right),\\
 \Delta_0=&-\frac{Q^2}{48m} -\frac{Q(\mu_3+\mu_3')}{4m}.
\end{split}
\ee
In the nonrelativistic approximation $\Delta_0 = 0$.

Note that from Eq.~(\ref{photon_momenta}) it follows that at $Q^2\ll 32 \langle\boldsymbol\mu^2\rangle$
$$\Delta_{1}^2\Delta_{2}^2 \approx \tfrac1{16}Q^2\left[Q^2 + \mathcal O(\mu^2)\right] $$ 
and the photon propagators can be moved out of the integral. For the wave functions obtained with both Paris \cite{Paris} and CD-Bonn \cite{cd-Bonn} potentials the estimated value of $\langle\boldsymbol\mu^2\rangle$ is $\sim 3\times 10^{-3}$~GeV$^2$ and we obtain the following restrictions on the applicability of our model
\be\label{restriction}
0.1\; \mathrm{GeV}^2\ll Q^2 \ll 16 m^2.
\ee

For rough estimation of the TPE effects we will use the nonrelativistic approximation for the nucleon electromagnetic current
\be\label{nonrel_current}
\begin{split}
&\langle \mathbf p',\xi',t_3|\Gamma_N^\mu\left(\Delta\right)|\mathbf p,\xi,t_3\rangle
\approx 2m\chi^\dag_{\xi'} \\
& \hspace{1.5cm}\times \left\{
\begin{array}{ll}
{{\widetilde\Gamma}_p ^\mu}\left(\Delta\right) \chi_{\xi}, & \text{if}\ t_3=\tfrac12,\\[0.25cm]
{{\widetilde\Gamma}_n ^\mu}\left(\Delta\right) \chi_{\xi}, & \text{if}\ t_3=-\tfrac12,
\end{array}
\right.
\end{split}
\ee
where
\be\label{nonrel_current1}
\begin{split}
{\widetilde\Gamma}_N ^0\left(\Delta\right) =& G_{EN}(-\Delta^2),\\
\widetilde{\boldsymbol\Gamma}_N\left(\Delta\right) =&\frac1{2m}\left\{
i(\boldsymbol \sigma \times \boldsymbol \Delta) G_{MN}(-\Delta^2) \right.\\
&\left.+
\left[-\boldsymbol \upsilon \pm(\boldsymbol \mu +\boldsymbol \mu')\right]F_{2N}(-\Delta^2)
\right\}  \\
\approx & \frac{i}{2m}(\boldsymbol \sigma \times \boldsymbol \Delta) G_{MN}(-\Delta^2).
\end{split}
\ee
We will also put $\Delta_1\approx \Delta_2\approx \tfrac12 q$ in the numerator of the electron propagator, while the denominator needs more care and we have to keep here terms proportional to $\boldsymbol \mu$ and $\boldsymbol \mu'$
\bee
&l^2=\ds\frac{\widetilde Q^2}4 + Q\left[\frac{Q}{4m}(\mu_3 + \mu'_3)\csc\frac{\theta}2
+(\mu_1 - \mu_1')\sin\frac{\theta}2\right],\nonumber\\
&\widetilde Q^2=Q^2\left(1 + \ds{\frac{Q}{12m}\csc\frac{\theta}2}\right).\label{denominator}
\eee

Now the amplitude becomes
\be\label{MII.b}
\Delta T^\mathrm{II(a)}_{h\sigma'\sigma}=\frac{24\pi\alpha}{\varepsilon Q^2}
\tau^h_{\mu\nu}\mathcal{S}_{\sigma'\sigma}^{\mathrm{(a)}\mu\nu},
\ee
where
\be\label{where2}
\begin{split}
 \mathcal{S}_{\sigma'\sigma}^{\mathrm{(a)}\mu\nu}= \int \frac{d^3\nu d^3\mu d^3\mu'\Psi^\dag_{\sigma'} (\boldsymbol \mu', \boldsymbol \nu') O^{\mu\nu} \Psi_\sigma (\boldsymbol \mu,\boldsymbol \nu)}{(2\pi)^3(l^2 +i0)},
\end{split}
\ee
$l^2$ is determined by Eq.~(\ref{denominator}) and 
$   O^{\mu\nu}= \widetilde{\Gamma}_{N_{2}}^\mu\left(\tfrac12 q\right)\widetilde{\Gamma}_{N_{3}}^\nu\left(\tfrac12 q\right)$.
\begin{figure*}
  \centering
  \includegraphics[height=0.2\textheight]{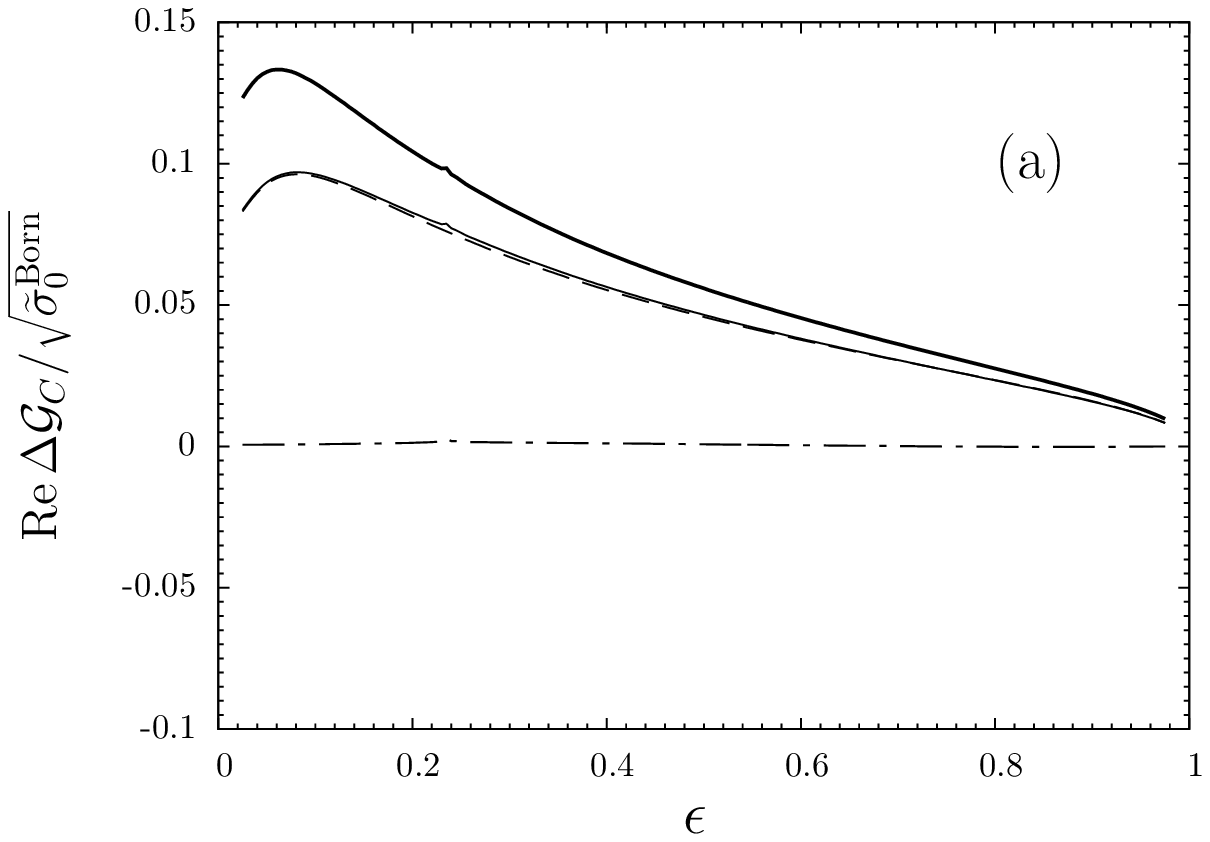}
 \hspace{1.cm} \includegraphics[height=0.2\textheight]{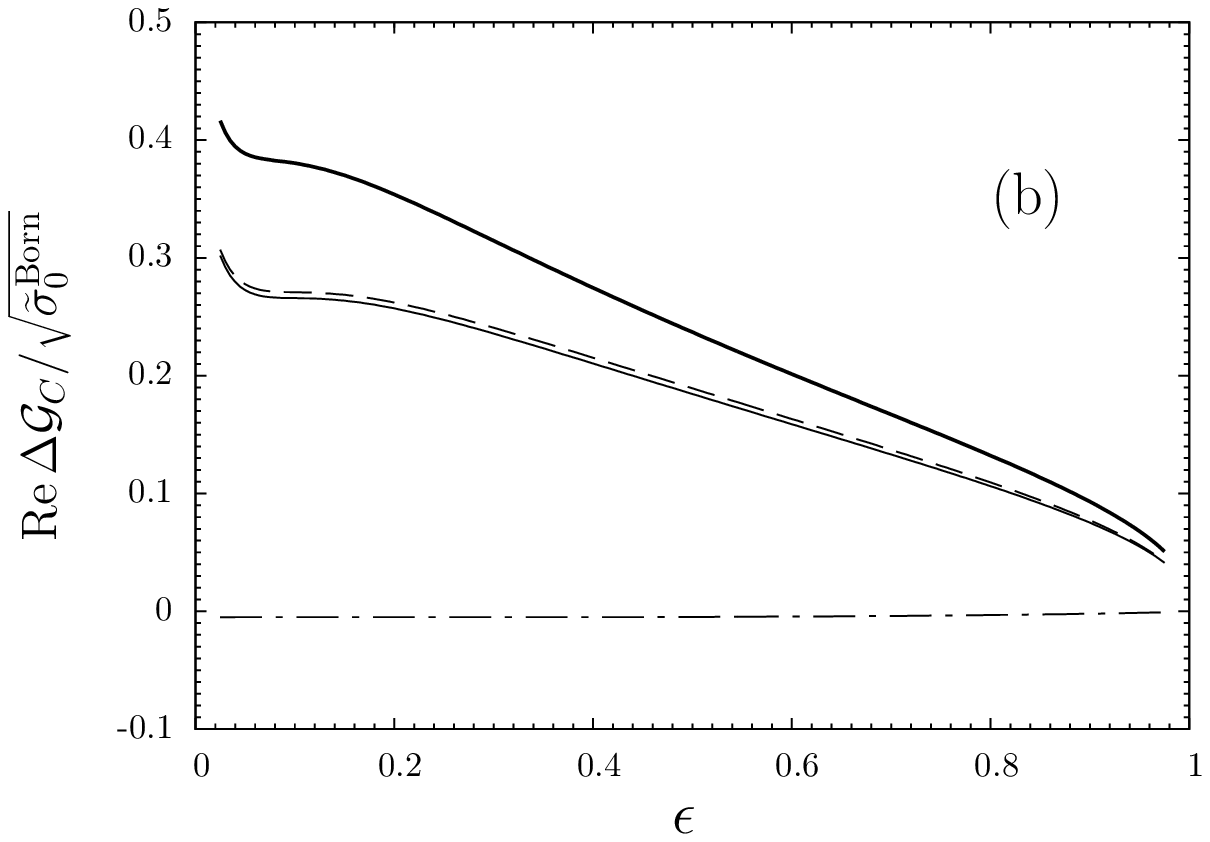}\\
  \includegraphics[height=0.2\textheight]{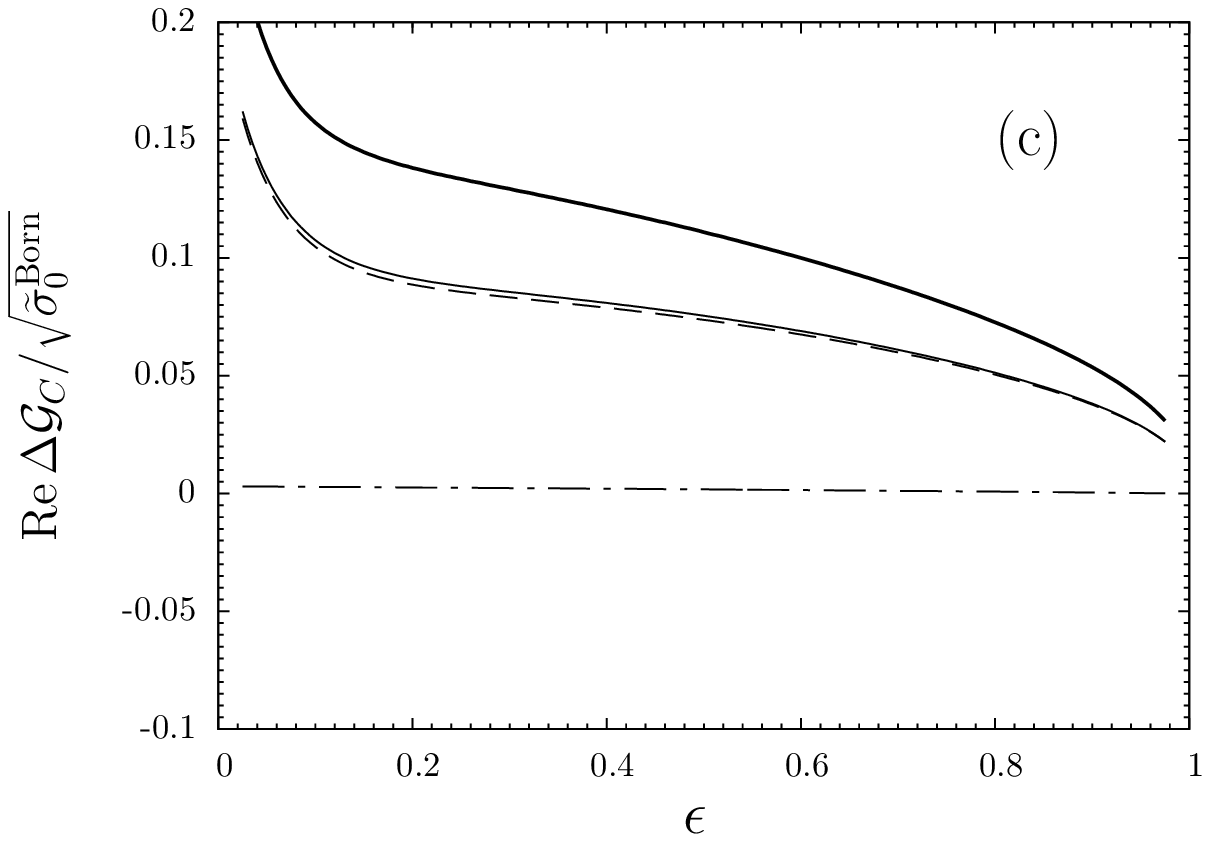}
 \hspace{1.cm} \includegraphics[height=0.2\textheight]{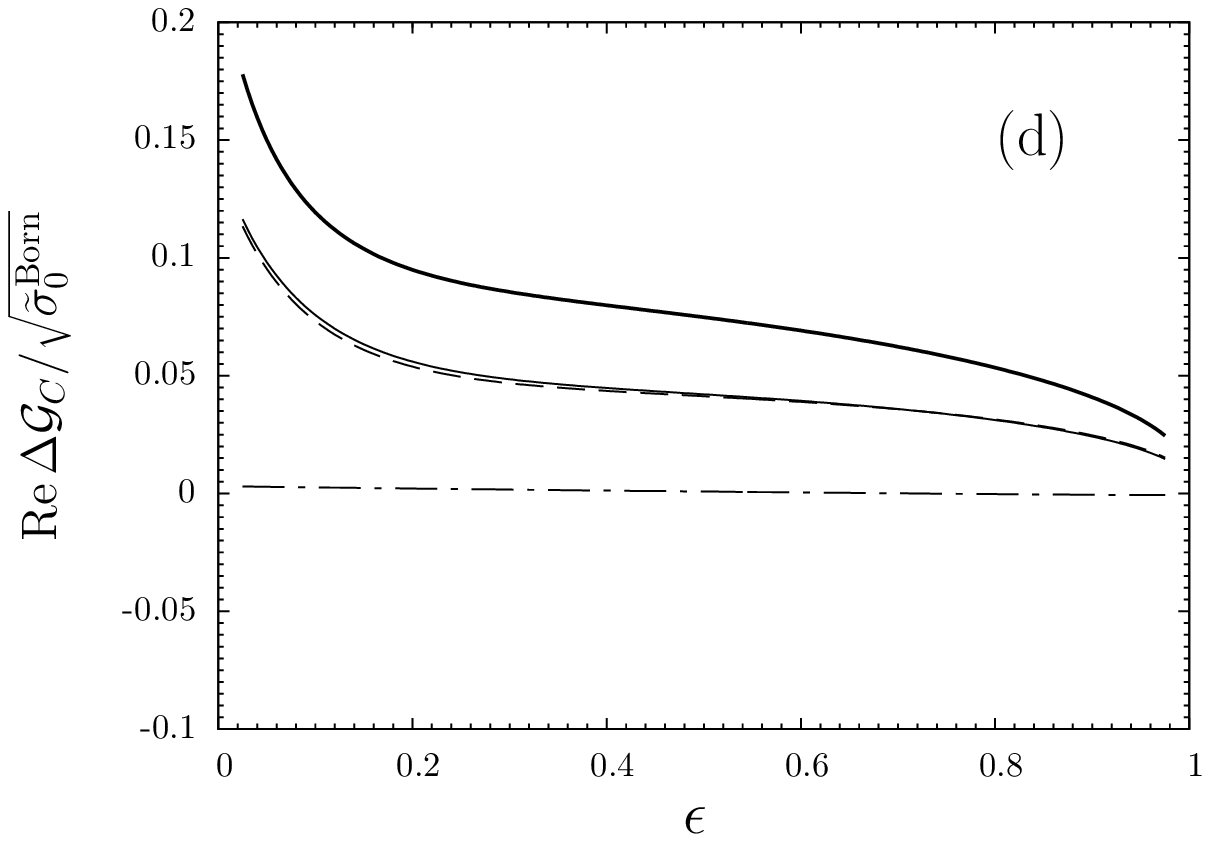}\\
 \includegraphics[height=0.2\textheight]{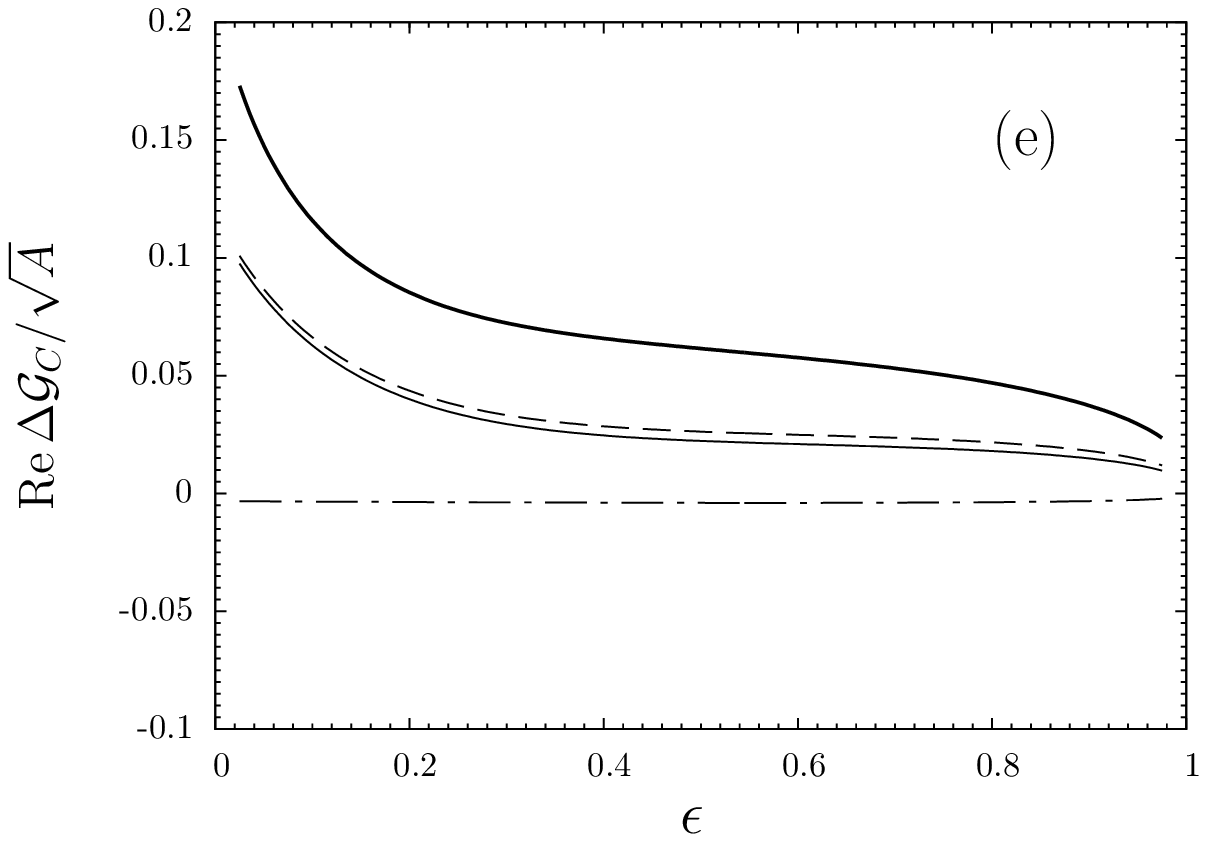}
\caption{The $\epsilon$ behavior of the ratio $\mathrm{Re}\Delta\mathcal G_E(Q^2,\epsilon)/\sqrt{\widetilde\sigma_0^\mathrm{Born}}$ at $Q^2=$5, 10, 20, 30, and 40 fm$^{-2}$ [(a), (b), (c), (d), and (e), respectively]. Dashed, dot-dashed, and solid (thin) curves are for $\mathcal M^{\mathrm I}$,  $\mathcal M^{\mathrm{II}}$, and $\mathcal M^{\mathrm I}+ \mathcal M^{\mathrm{II}}$, respectively, calculated for the Paris potential. The solid (bold) curves depict $\mathcal M^{\mathrm I} + \mathcal M^{\mathrm{II}}$, calculated for the CD-Bonn potential.
}
\label{fig:dGC}
\end{figure*}
\begin{figure*}
  \centering
  \includegraphics[height=0.2\textheight]{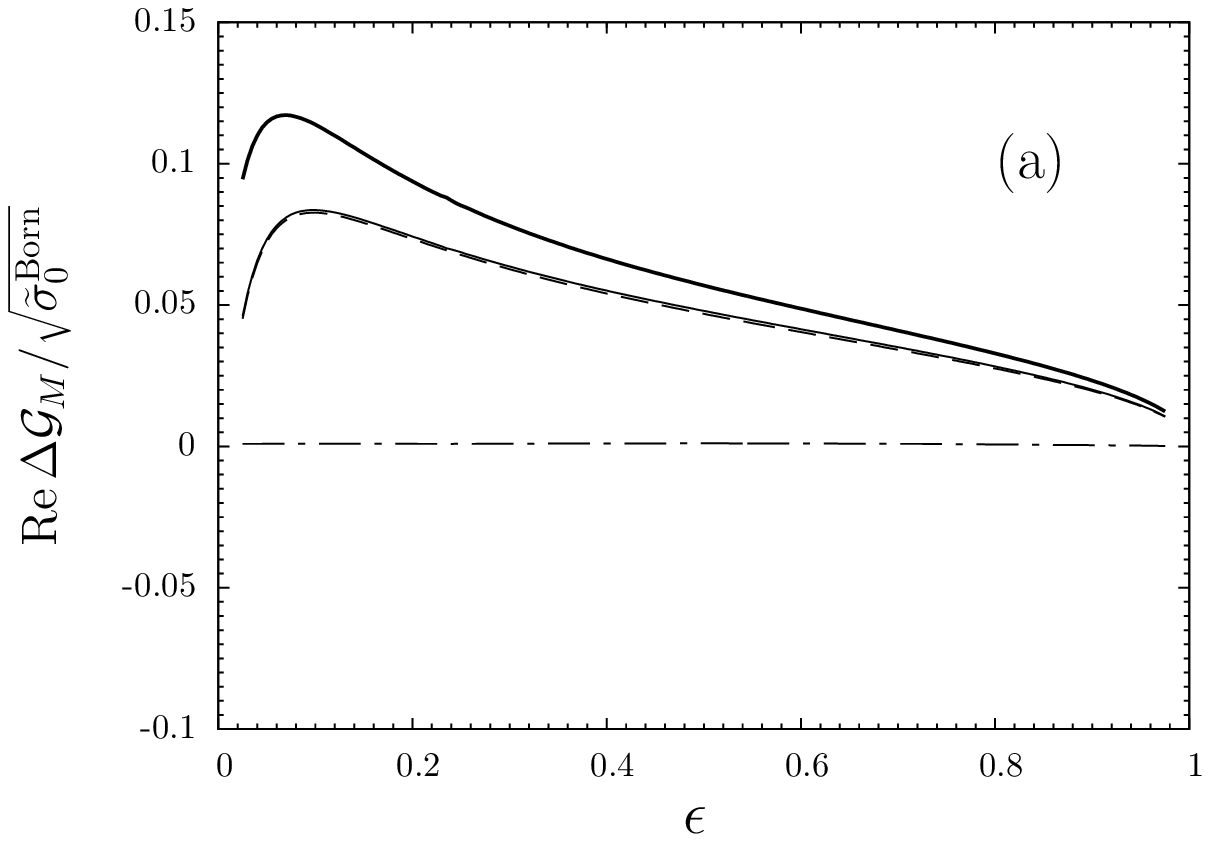}
 \hspace{1.cm} \includegraphics[height=0.2\textheight]{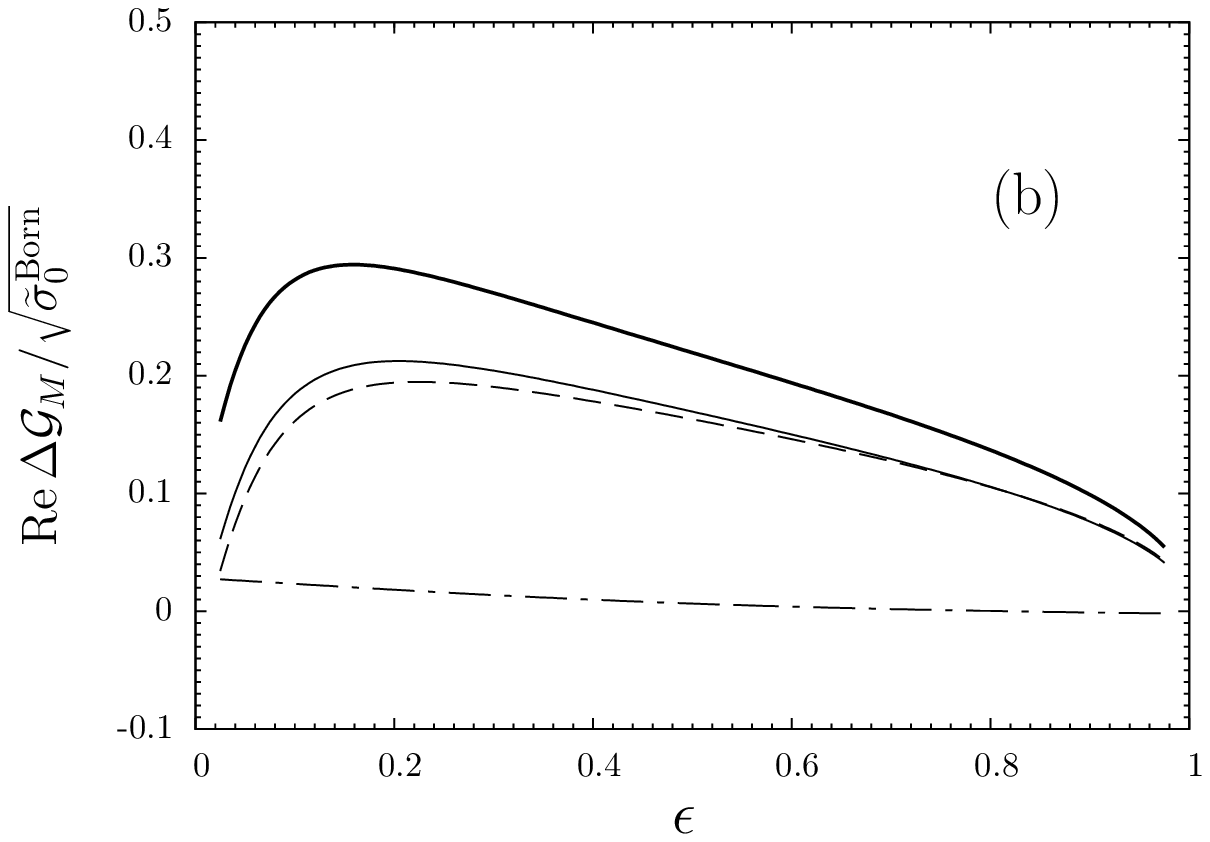}\\
  \includegraphics[height=0.2\textheight]{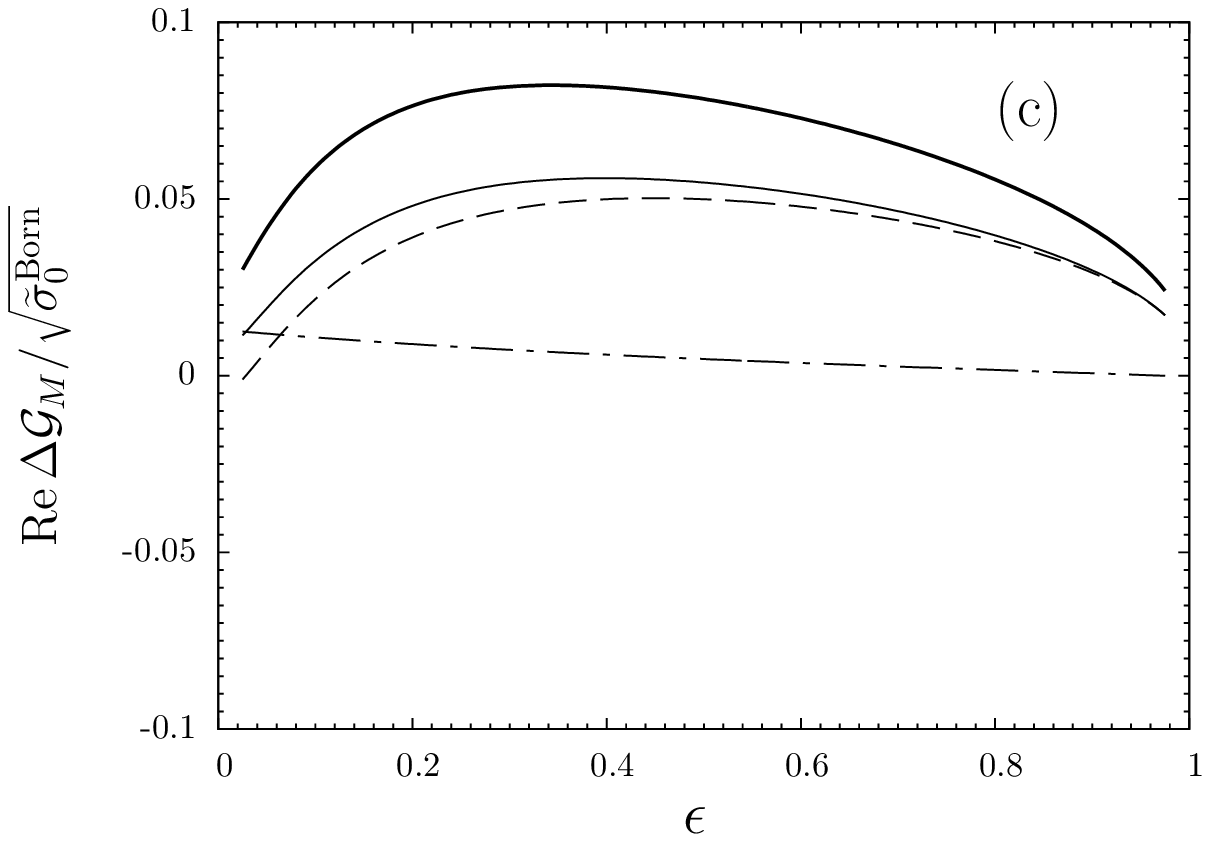}
 \hspace{1.cm} \includegraphics[height=0.2\textheight]{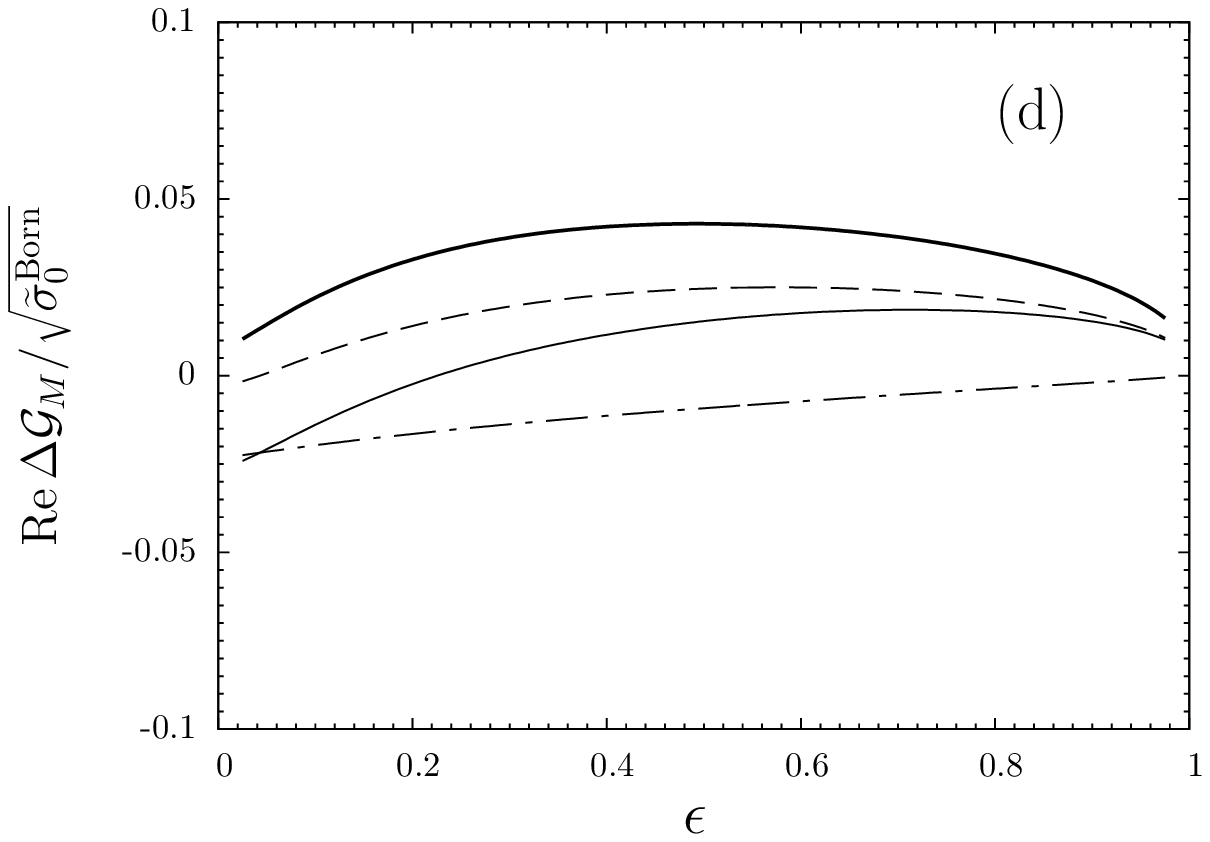}\\
 \includegraphics[height=0.2\textheight]{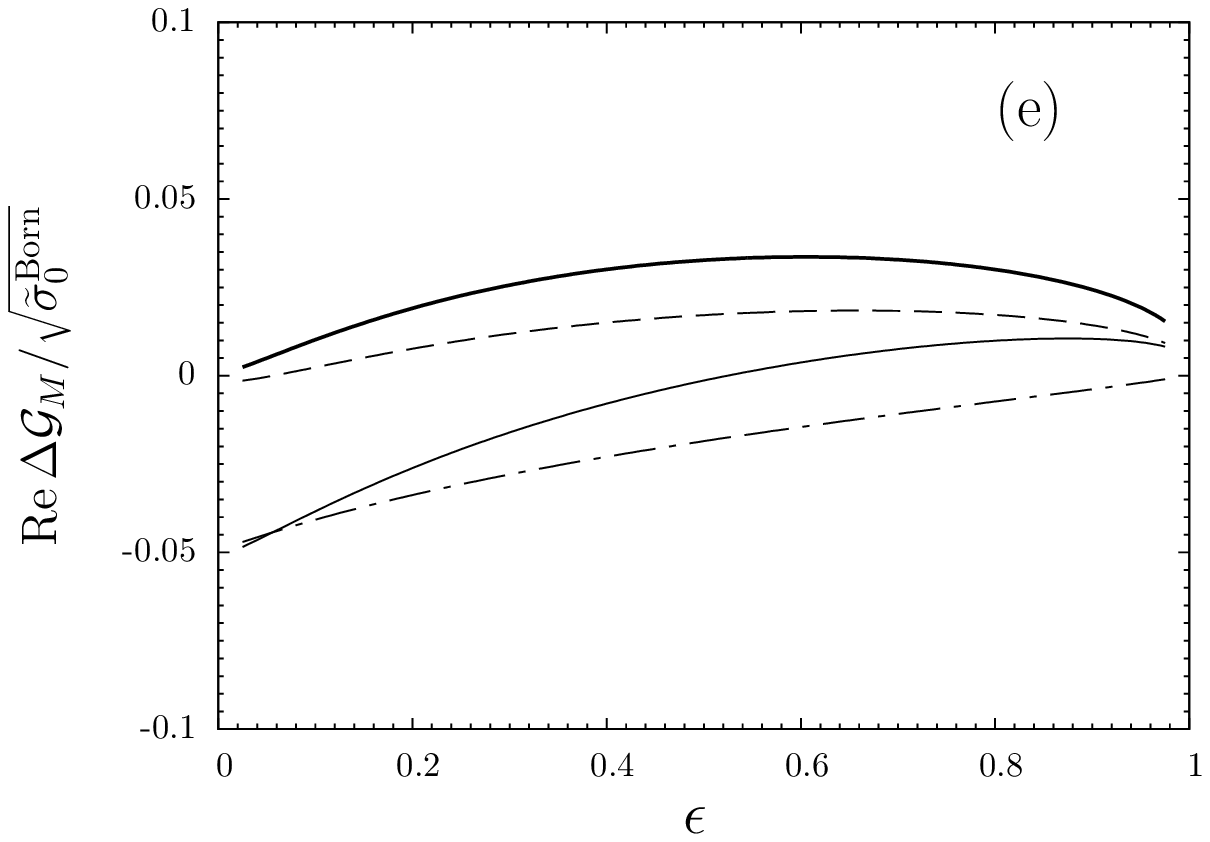}
\caption{The same as Fig.~\ref{fig:dGC} for the ratio $\mathrm{Re}\Delta\mathcal G_M(Q^2,\epsilon)/\sqrt{\widetilde\sigma_0^\mathrm{Born}}$.
}
\label{fig:dGM}
\end{figure*}
\begin{figure*}
  \centering
  \includegraphics[height=0.2\textheight]{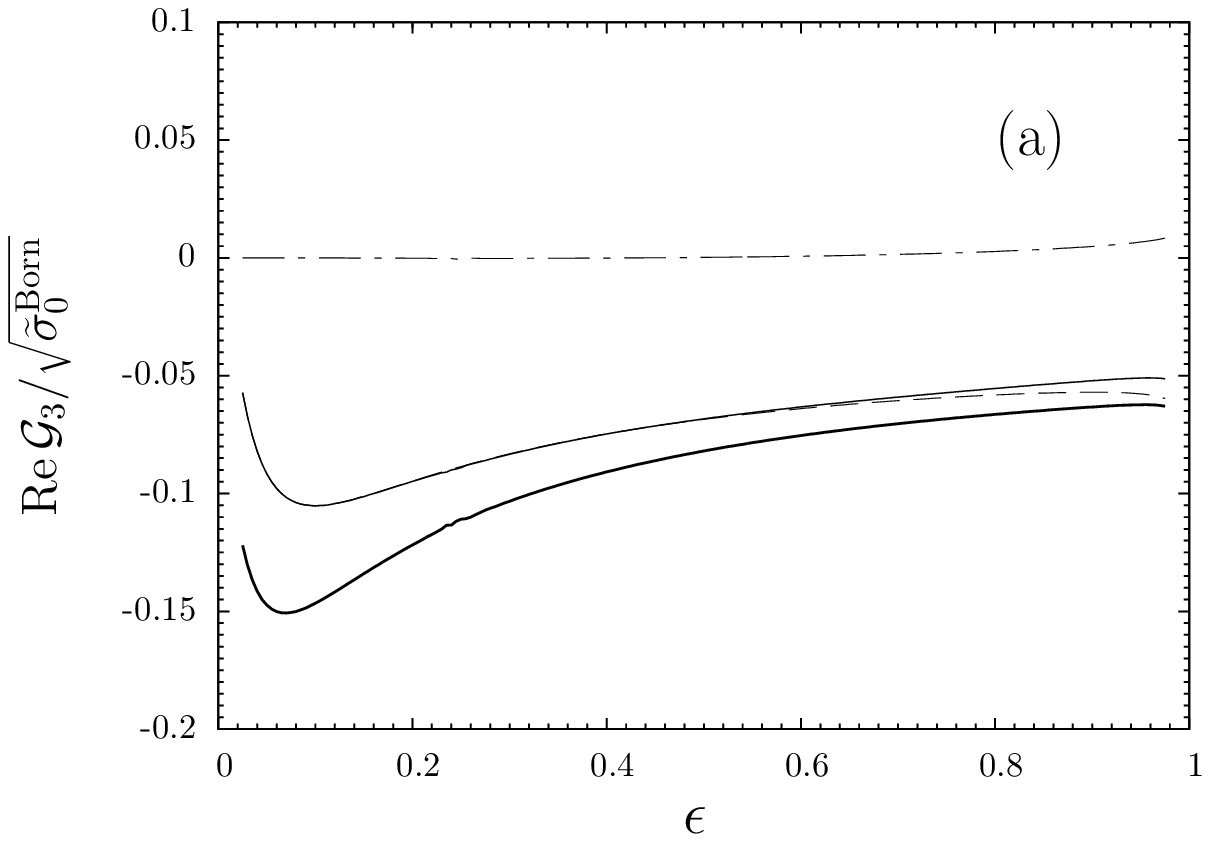}
 \hspace{1.cm} \includegraphics[height=0.2\textheight]{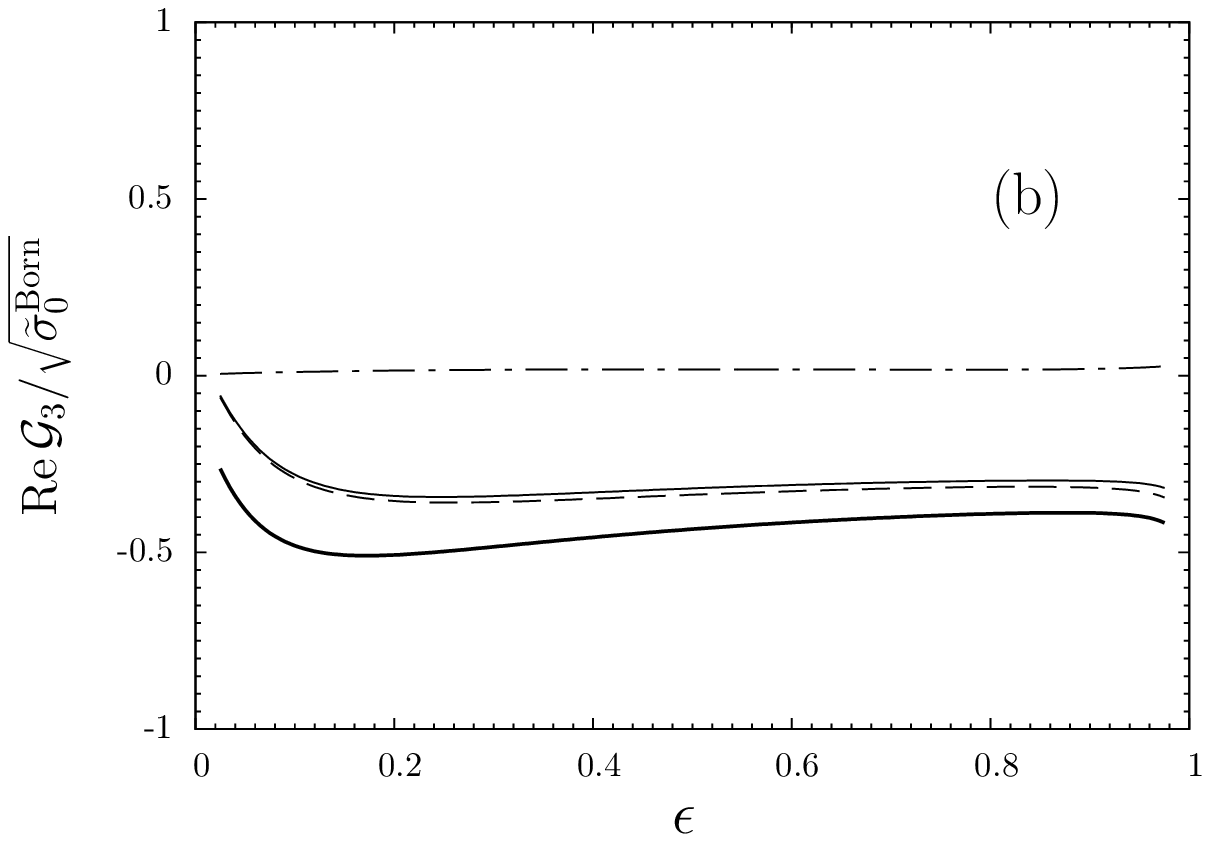}\\
  \includegraphics[height=0.2\textheight]{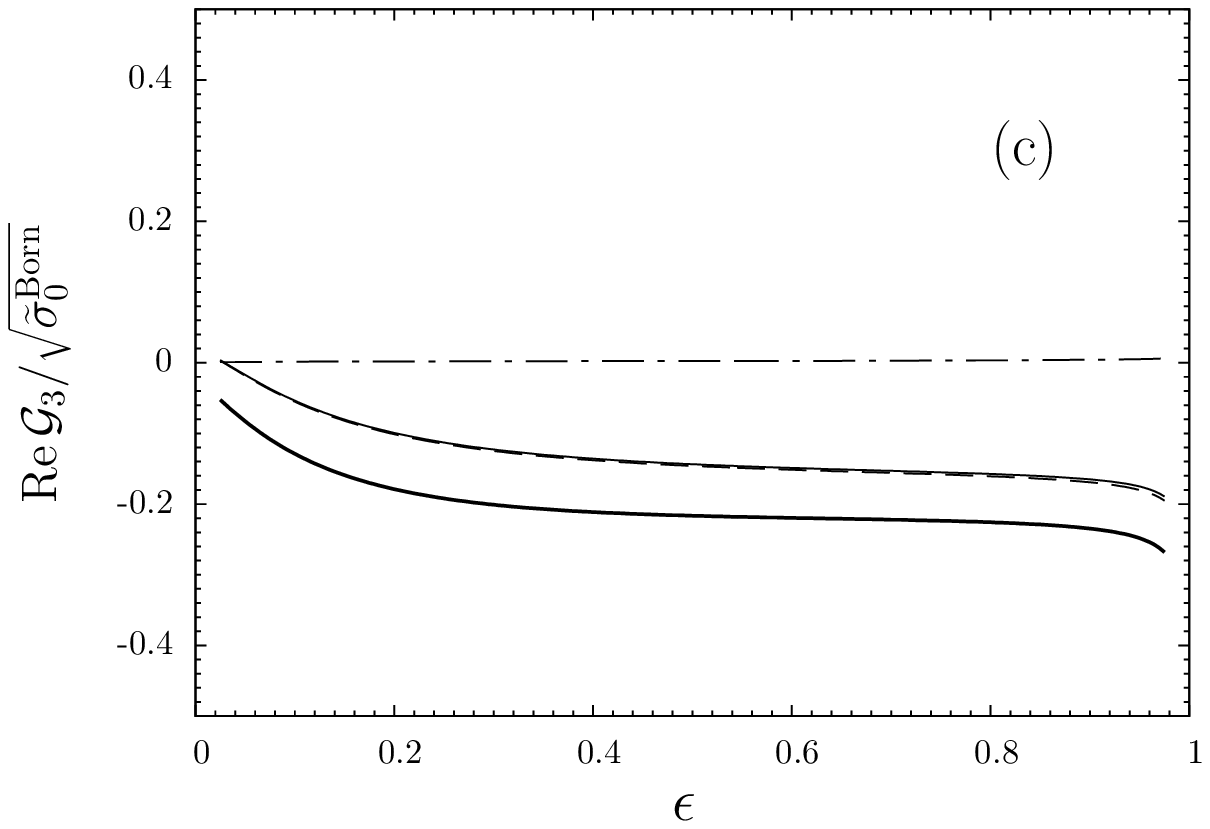}
 \hspace{1.cm} \includegraphics[height=0.2\textheight]{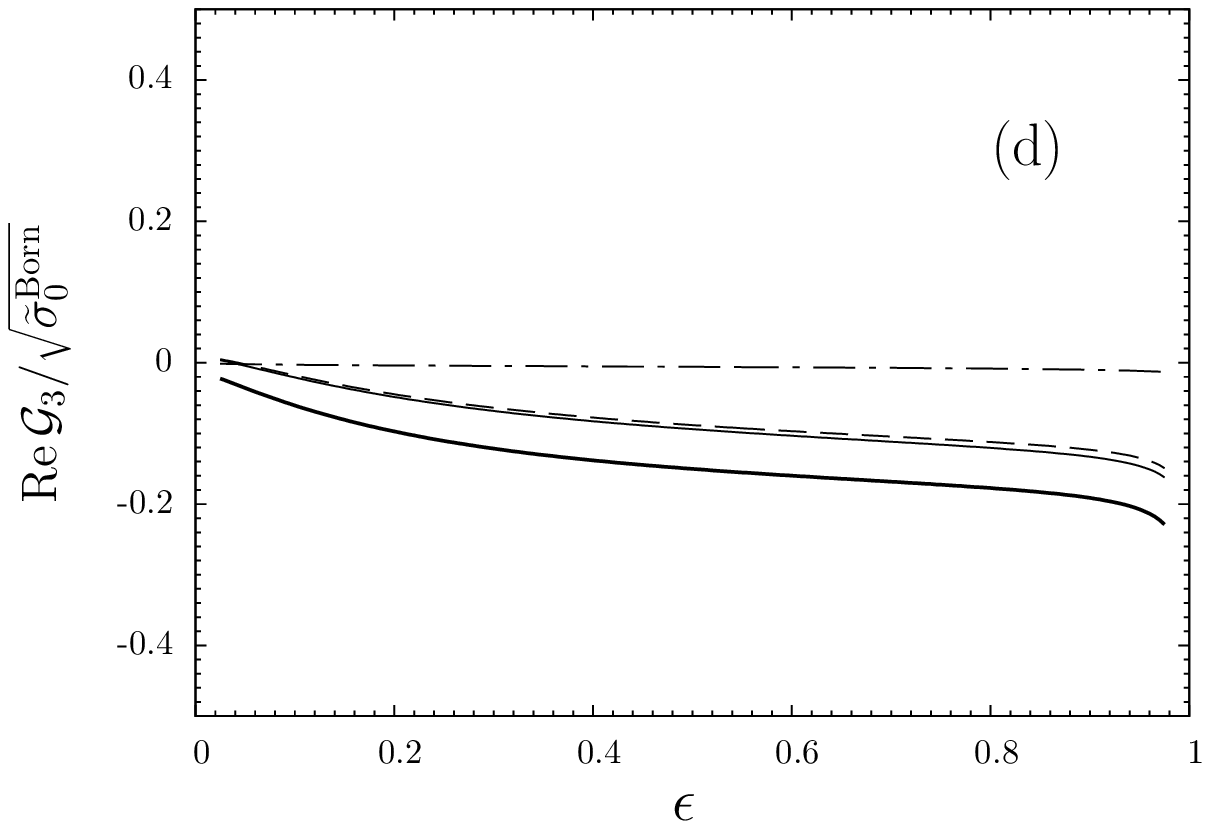}\\
 \includegraphics[height=0.2\textheight]{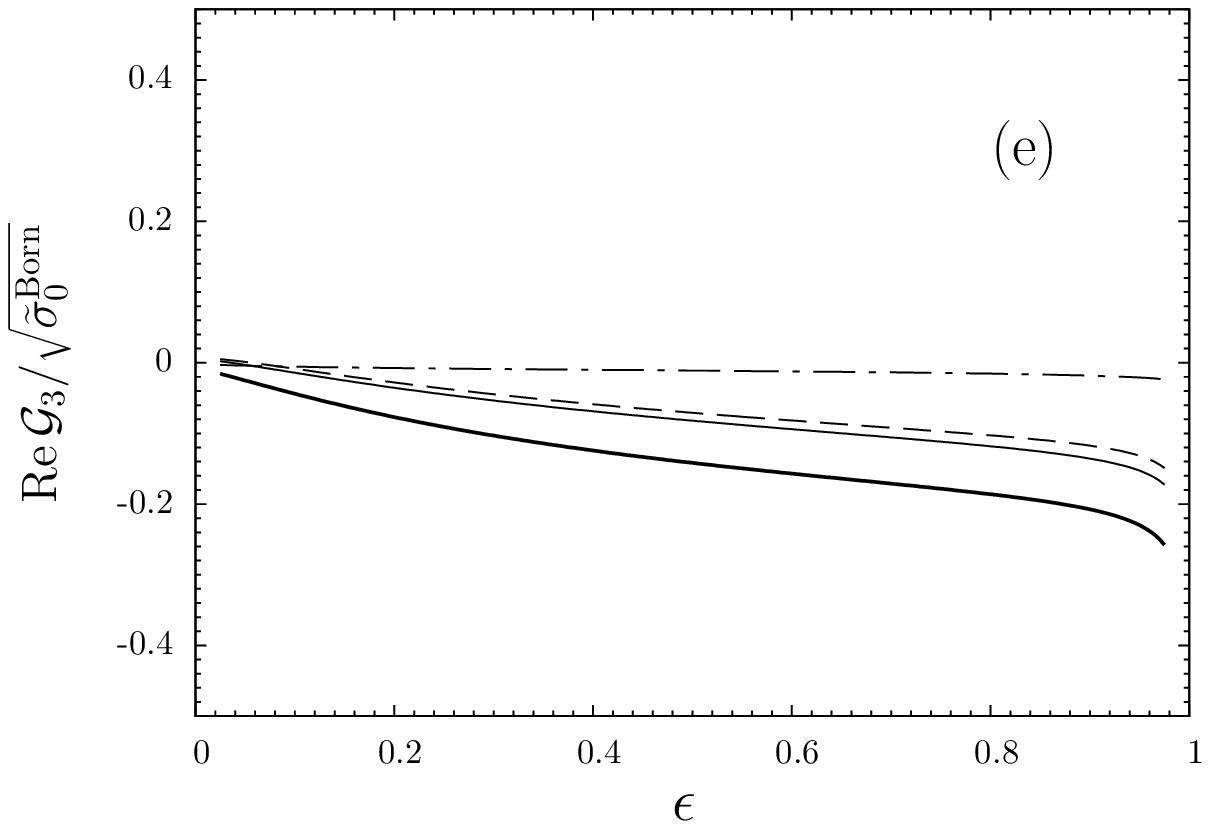}
\caption{The same as Fig.~\ref{fig:dGC} for the ratio $\mathrm{Re}\mathcal F_3(Q^2,\epsilon)/\sqrt{\widetilde\sigma_0^\mathrm{Born}}$, where $A(Q^2)=G_E^2(Q^2)+Z^{-2}\left(\frac{\mu M}{m}\right)^2 \eta G^2_M(Q^2)
$.
}
\label{fig:F3}
\end{figure*}
Using the integral representation
\be\label{IR}
\dfrac{1}{a +i0}=-i\int_0^\infty dt e^{ia t},
\ee
and performing the Fourier transform of the wave functions we reduce the 9-dimensional integral (\ref{where2}) to the following 4-dimensional integral
\be\label{calculations1}
\begin{split}
& \mathcal{S}_{\sigma'\sigma}^{\mathrm{(a)} \mu\nu}
= -i\frac{\sin\tfrac{\theta}2}Q  \int_0^\infty dt e^{\frac{i}4tQ\left(\sin\frac{\theta}2 +\frac{Q}{12m}\right)} \\
&\times \int d^3\rho e^{\frac{i}3\mathbf q\cdot \boldsymbol \rho}
 \Phi^\dag_{\sigma'} (t\mathbf y',\boldsymbol \rho)O^{\mu\nu}\Phi_\sigma (t\mathbf y,\boldsymbol \rho)\\
 &=-\frac{i\sin\tfrac{\theta}2}{y^2Q}  \int_0^\infty \frac{dt}{t^2} e^{\frac{i}4tQ\left(\sin\frac{\theta}2 +\frac{Q}{12m}\right)}\mathfrak{M}_{\sigma'\sigma}^{\mu\nu},
\end{split}
\ee
where
\be
\begin{split}
 &\mathbf y=\left(\cos\frac{\theta}2,0,\frac{Q}{4m}\right)=y(\sin\Theta,0,\cos\Theta),\\
 &\mathbf y'=\left(\cos\frac{\theta}2,0,-\frac{Q}{4m}\right)=y(\sin\Theta,0,-\cos\Theta),\\
 &y=\sqrt{\cos^2\frac{\theta}2+\frac{Q^2}{16m^2}},\qquad \sin\Theta=\frac{\cos\frac{\theta}2}{y}.
\end{split}
\ee

From similar calculations we get that equations, identical to Eq.~(\ref{MII.b}), have place for all $\Delta T^\mathrm{II(i)}_{h\sigma'\sigma}$ (i=a,b,c,d)
\be\label{similar_Eqs}
T^\mathrm{II(i)}_{h\sigma'\sigma}=\frac{24\pi\alpha}{\varepsilon Q^2}
\tau^h_{\mu\nu}\mathcal{S}_{\sigma'\sigma}^{\mathrm{(i)}\mu\nu}
\ee
with
$ \mathcal{S}_{\sigma'\sigma}^{\mathrm{(b)}\mu\nu} = \mathcal{S}_{\sigma'\sigma}^{\mathrm{(a)}\mu\nu} $ and
\be\label{calculations}
\begin{split}
& \mathcal{S}_{\sigma'\sigma}^{\mathrm{(c)}\mu\nu}= \mathcal{S}_{\sigma'\sigma}^{\mathrm{(d)}\mu\nu}\\
 &=-i \frac{\sin\tfrac{\theta}2}{y^2Q}  \int_0^\infty \frac{dt}{t^2} e^{\frac{i}4tQ\sin\frac{\theta}2}\mathfrak{M}_{\sigma'\sigma}^{\mu\nu}.
\end{split}
\ee
and thus the total $\Delta T^\mathrm{II}_{h\sigma'\sigma}=\frac{24\pi\alpha}{\varepsilon Q^2}
\tau^h_{\mu\nu}\mathcal{S}_{\sigma'\sigma}^{\mu\nu}$, where
\be\label{S-factor}
\mathcal{S}_{\sigma'\sigma}^{\mu\nu}=2\left(\mathcal{S}_{\sigma'\sigma}^{\mathrm{(a)}\mu\nu} + \mathcal{S}_{\sigma'\sigma}^{\mathrm{(c)}\mu\nu}\right).
\ee

The non-vanishing components of the tensor $\mathfrak{M}_{\sigma'\sigma}^{\mu\nu}$ are
\be\label{non-vanishing}
\begin{split}
 & \mathfrak{M}_{\sigma'\sigma}^{    00}=\delta_{\sigma'\sigma}\mathfrak{M}^{00},\\
 & \mathfrak{M}_{\sigma'\sigma}^{  0\pm}=\pm\delta_{\sigma',\sigma\pm 1}\mathfrak{M}^{0+},\\
 & \mathfrak{M}_{\sigma'\sigma}^{\pm\mp}=\delta_{\sigma'\sigma}\mathfrak{M}^{+-}.
\end{split}
\ee

From numerical calculations we find that contribution of the wave function components with $l=2$ is less than 10\% and in further estimations of the TPE effects we ignore these components. A similar situation takes place in the $ed$-scattering where the $D$ wave gives small contribution to the TPE effects \cite{KK-ED}. The reason is very simple: the main contribution in the integral of Eq.~(\ref{calculations}) comes from small $r$, where the components with $l=2$ are suppressed by a factor of $r^2$.

The quantities $\mathfrak{M}^{00}$, $\mathfrak{M}^{0+}$, and $\mathfrak{M}^{+-}$ are given in Appendix~\ref{App1}.

Taking into account explicit expression for the lepton tensor
\begin{equation}
\label{LeptonComps}
\begin{split}
&\tau^h_{00}=8\varepsilon^2\cos\textstyle{\frac{\theta}2}, \qquad \tau^h_{-+}=4 \varepsilon^2\cos\textstyle{\frac{\theta}2},\\
&\tau^h_{0\pm}=2\sqrt{2} \varepsilon^2\left(1 + \cos^2\textstyle{\frac{\theta}2}\mp h\sin\textstyle{\frac{\theta}2}\right),
\end{split}
\end{equation}
we arrive at the final expression for the correction to the reduced amplitude
\begin{widetext}
\be\label{TII}
\Delta T^\mathrm{II}_h = \frac{192 \pi\alpha \varepsilon}{Q^2}\left(
\begin{array}{ll}
\cos\tfrac{\theta}2 (\mathcal{S}^{00}+\mathcal{S}^{+-} ) & -\frac{\mathcal{S}^{0+}}{2\sqrt2} (1 + \cos^2\tfrac{\theta}2-h\sin\tfrac{\theta}2)\\[0.25cm]
-\frac{\mathcal{S}^{0+}}{2\sqrt2}(1 + \cos^2\tfrac{\theta}2+h\sin\tfrac{\theta}2) & \cos\tfrac{\theta}2 (\mathcal{S}^{00}+\mathcal{S}^{+-} )
\end{array}
\right),
\ee
\end{widetext}
where $\mathcal{S}^{00}$, $\mathcal{S}^{0+}$, and $\mathcal{S}^{+-}$ are defined by equations similar to Eqs.~(\ref{non-vanishing})
\be\label{non-vanishing.1}
\begin{split}
 & \mathcal{S}_{\sigma'\sigma}^{ 00}=\delta_{\sigma'\sigma}\mathcal{S}^{00},\\
 & \mathcal{S}_{\sigma'\sigma}^{0\pm}=\pm\delta_{\sigma',\sigma\pm 1}\mathcal{S}^{0+},\\
 & \mathcal{S}_{\sigma'\sigma}^{\pm\mp}=\delta_{\sigma'\sigma}\mathcal{S}^{+-}.
\end{split}
\ee

Comparing this result with Eq.~(\ref{spin_structure}) 
we get
\be\label{TII.final}
\begin{split}
\Delta\mathcal G_E^\mathrm{II}&=\frac{192\pi\alpha \varepsilon}{Q^2}Z^{-1}\left(\mathcal{S}^{00}+\mathcal{S}^{+-}\right),\\
\Delta\mathcal G_M^\mathrm{II}&=\frac{96\pi\alpha \varepsilon m}{\mu MQ^2\sqrt{2\eta}}
\left(1 + \epsilon\right)\mathcal{S}^{0+},\\
\mathcal F_3^\mathrm{II}&=\frac{96\pi\alpha  M^2}{EQ^2\sqrt{2\eta}}\ \mathcal{S}^{0+}.
\end{split}
\ee
\section{Numerical results and conclusions\label{sec:results}}
Figures \ref{fig:dGC}, \ref{fig:dGM}, and \ref{fig:F3} display the $\epsilon$ behavior of the ratios 
$\mathrm{Re}\Delta\mathcal G_C(Q^2,\epsilon)/\sqrt{\widetilde\sigma_0^\mathrm{Born}}$, 
$\mathrm{Re}\Delta\mathcal G_M(Q^2,\epsilon)/\sqrt{\widetilde\sigma_0^\mathrm{Born}}$, 
and 
$\mathrm{Re}\mathcal G_3(Q^2,\epsilon)/\sqrt{\widetilde\sigma_0^\mathrm{Born}}$ at fixed $Q^2$.
Here
\be
 \widetilde\sigma_0^\mathrm{Born}=0.5 G_C^2(Q^2)+\eta\left(\frac{\mu(^3\mathrm{He})M}m\right)^2G_M^2(Q^2)
\ee
is the Born approximation for the reduced cross section at $\epsilon=0.5$.
Two $Q^2$ values (10 and 20 fm$^{-2}$) are chosen in the vicinity of the region where $G_C$ and $G_M$ change sign. The form factors $G_C(Q^2)$ and $G_M(Q^2)$ were estimated from the parametrization of Ref.~\cite{Amroun}.

The nucleon TPE amplitudes, which are needed for the evaluation of the $\mathcal M^{\mathrm I}$ amplitude, were calculated with the {\tt TPEcalc} program \cite{TPEcalc} based on the dispersion method of Ref.~\cite{BK_PRC_2008}. For the magnetic form factors of the proton and neutron we use usual dipole parametrization;
the JLab parametrization, see Eq.~(33) of Ref.~\cite{PPV_2007} 
and so-called Galster parametrization \cite{Galster} are used for the electric form factors of the proton and neutron, respectively.

The numerical calculations are done with the trinucleon wave functions for the Paris \cite{Paris} and CD-Bonn \cite{cd-Bonn} potentials.

\begin{figure*}
  \centering
  \includegraphics[height=0.2\textheight]{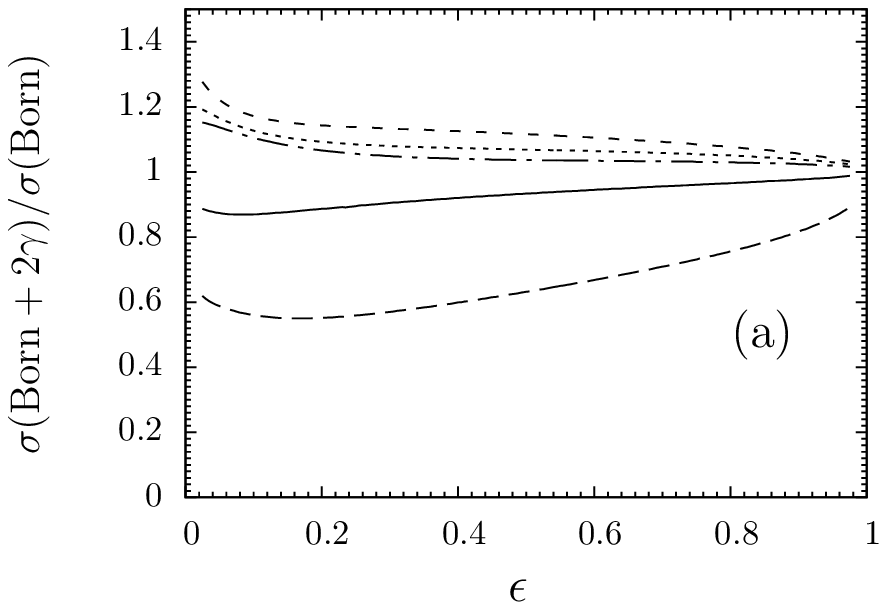}
 \hspace{1.cm} \includegraphics[height=0.2\textheight]{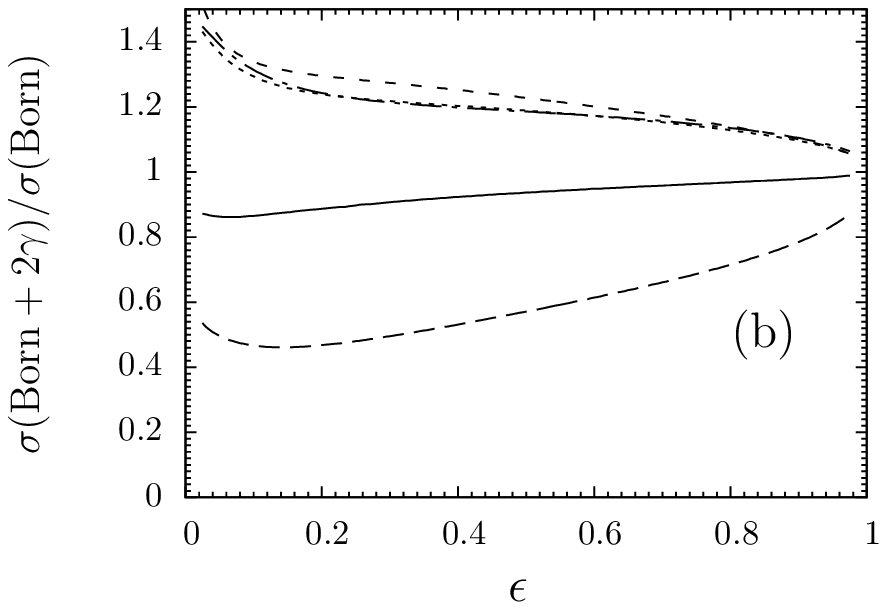}
\caption{The $\epsilon$ dependence of the ratio of the differential cross section calculated in the Born+TPE approximation to the Born approximation, at fixed $Q^2$. The $^3$He wave functions for the Paris (a) and CD-Bonn (b) potentials. The solid, dashed, short-dashed, dotted and dash-dotted lines are for $Q^2=$ 5, 10, 20, 30, and 40 fm$^{-2}$, respectively.
}
\label{fig:cs_ratio}
\end{figure*}

In Fig.~\ref{fig:cs_ratio} we show the $\epsilon$ dependence of the ratio of the differential cross section, calculated in the Born+TPE approximation, to the Born approximation, at fixed $Q^2$.

From the results of our numerical calculations we drew the following conclusions:
\begin{itemize}
 \item The TPE corrections are a few times more significant in the scattering of electrons off $^3$He than in the electron-nucleon scattering.
 \item The contribution of the type I diagram to the generalized electric form factor is minor at all values of $Q^2$, while in the generalized magnetic form factor its contribution increases with $Q^2$ and becomes significant at $Q^2\gtrsim$ 30 fm$^{-2}$.
 \item The $\mathcal F_3$ form factor is large and it would be interesting to observe it experimentally.
 \item Similarly to the electron-deuteron scattering \cite{KK-ED} the main source of uncertainty in the estimated TPE corrections comes from the short-range part of the trinucleon wave function.
\end{itemize}

\acknowledgments
The authors acknowledge Dmitry Borisyuk for fruitful discussions on various points considered in the present paper and reading the manuscript.

\appendix
\section{Connection between wave function and vertex function\label{App2}}
To establish connection between the trinucleon wave function with two nucleons on mass shell and the vertex function $V(p_1,p_2,p_3)$ let us consider the electromagnetic current of the trinucleon in the impulse approximation. Due to antisymmetry of the trinucleon wave function one may calculate only the diagram where a virtual photon interacts with the first nucleon and then multiply the result by the number of nucleons,
\be\label{IA.1}
\begin{split}
 & i\left<\mathbf P'\sigma'\right|J_\mu\left|\mathbf P\sigma\right>=
   3\int\frac{d^4\mu d^4\nu}{(2\pi)^8}\left<\mathbf P'\sigma'\right|\\
 & \times iV^\dag (p'_1,p_2,p_3)\frac{i(p\!\!\!/'_1+m)i\Gamma_{N_1\mu} i(p\!\!\!/_1+m)}{({p'_1}^2-m^2+i0)({p_1}^2-m^2+i0)}\\
 & \times \frac{i(p\!\!\!/_2+m)}{{p_2}^2-m^2+i0}\cdot \frac{i(p\!\!\!/_3+m)}{p_3^2-m^2+i0}iV (p_1,p_2,p_3)
   \left|\mathbf P\sigma\right>\\
 & =3i\int\frac{d^3\mu d^3\nu}{2E_22E_3(2\pi)^6} \left<\mathbf P'\sigma'\right|V^\dag (p'_1,p_2,p_3)\\
 & \times  \frac{p\!\!\!/'_1+m}{{p'_1}^2-m^2+i0}\Gamma_{N_1\mu}\frac{p\!\!\!/_1+m}{{p_1}^2-m^2+i0}\\
 & \times (p\!\!\!/_2+m)(p\!\!\!/_3+m)V (p_1,p_2,p_3)
   \left|\mathbf P\sigma\right> . 
\end{split}
\ee
In the nonrelativistic approximation this becomes
\be\label{IA.2}
\begin{split}
 & \left<\mathbf P'\sigma'\right|J_\mu\left|\mathbf P\sigma\right>=
 3 (2m)^2\int\frac{d^3\mu d^3\nu}{(2\pi)^6} \\
 & \times \left<\mathbf P'\sigma'\right| \frac{V^\dag (p'_1,p_2,p_3)}{{p'_1}^2-m^2+i0}\widetilde\Gamma_{N_1\mu}\frac{V (p_1,p_2,p_3)}{{p_1}^2-m^2+i0}
   \left|\mathbf P\sigma\right> , 
\end{split}
\ee
where $\widetilde\Gamma^{N_1}_\mu$ is determined by Eqs.~(\ref{nonrel_current1}). Equation~(\ref{IA.2}) must be compared with
\be\label{IA.3}
\begin{split}
 & \left<\mathbf P'\sigma'\right|J_\mu\left|\mathbf P\sigma\right>=
   6M \int d^3\mu d^3\nu \\
 & \times \Psi^\dag_{\sigma'} (\boldsymbol \mu,\boldsymbol \nu + \tfrac23 \mathbf q) \widetilde\Gamma_{N_1\mu}\Psi_{\sigma} (\boldsymbol \mu,\boldsymbol \nu),
\end{split}
\ee
from which we get
\be\label{IA.4}
\Psi_{\sigma} (\boldsymbol \mu,\boldsymbol \nu) = \frac{1}{(2\pi)^3}\frac{2m}{\sqrt{2M}}\frac{V(p_1,p_2,p_3)}{{p_1}^2-m^2 +i0} \left|\mathbf P,\sigma\right>.
\ee
\section{Calculation of the matrix element $\mathfrak{M}_{\sigma'\sigma\mu\nu}$\label{App1}}
To calculate this matrix element one needs the following matrix elements for the $N_2+N_3$ subsystem of the trinucleon 
\be\label{susystem_ME}
\begin{split}
& \left\langle 1 s_3;0 0 \right|\widetilde\Gamma_{N_2 0}\widetilde\Gamma_dGC_Q2{N_3 0}\left| 1s_3;00 \right\rangle =G_{EE}\\
& \sum_{\tau_3 t_3}\left\langle 1\tfrac12 \tau_3 t_3|\tfrac12\tfrac12\right\rangle^2
\left\langle 00;1\tau_3 \right|\widetilde\Gamma_{N_2 0}\widetilde\Gamma_{N_3 0}\left| 00;1\tau_3 \right\rangle\\
& \hspace{0.5cm}= G'_{EE},\\
& \left\langle 00;1 0 \right|\widetilde\Gamma_{N_2 0}\widetilde\Gamma_{N_3 0}\left| 10;00 \right\rangle =0,\\
& \left\langle 1 0;0 0 \right|\widetilde\Gamma_{N_2 0}\widetilde\Gamma_{N_3 +}\left| 1-1;00 \right\rangle = \left\langle 1 1;0 0 \right|\widetilde\Gamma_0^{(2)}\widetilde\Gamma_+^{(3)}\left| 10;00 \right\rangle\\
& \hspace{0.5cm}=
\sqrt{\tau} G_{EM},\\
& \left\langle 1\tfrac12 0 \tfrac12|\tfrac12\tfrac12\right\rangle
\left\langle 00;10 \right|\widetilde\Gamma_{N_2 0}\widetilde\Gamma_{N_3 +}\left| 1-1;00 \right\rangle\\
& =\sqrt{\tfrac{\tau}{3}} G'_{EM},\\
& \left\langle 1s_3;00 \right|\widetilde\Gamma_{N_2 +}\widetilde\Gamma_{N_3 -}\left| 1s_3;00 \right\rangle= -\delta_{s_3 0}\tau G_{MM}\\
& \sum_{\tau_3 t_3}\left\langle 1\tfrac12 \tau_3 t_3|\tfrac12\tfrac12\right\rangle^2
\left\langle 00;1\tau_3 \right|\widetilde\Gamma_{N_2 +}\widetilde\Gamma_{N_3 -}\left| 00;1\tau_3 \right\rangle\\
& \hspace{0.5cm}=- \tau G'_{MM},\\
& \left\langle 00;10 \right|\widetilde\Gamma_{N_2 +}\widetilde\Gamma_{N_3 -}\left| 10;00 \right\rangle=0,
\end{split}
\ee
where
\\[-1.cm]
\be\label{GEE_etc}
\begin{split}
 G_{EE}=& G_{Ep}\left(\tfrac14Q^2\right)G_{En}\left(\tfrac14Q^2\right) \\
 G'_{EE}=&\tfrac23 G_{Ep}^2\left(\tfrac14Q^2\right) \\
 & + \tfrac13 G_{Ep}\left(\tfrac14Q^2\right)G_{En}\left(\tfrac14Q^2\right),\\
 G_{EM}=&\tfrac12\left[G_{Ep}\left(\tfrac14Q^2\right)G_{Mn}\left(\tfrac14Q^2\right) \right.\\
 &\left.+ G_{En}\left(\tfrac14Q^2\right)G_{Mp}\left(\tfrac14Q^2\right)\right],\\
 G'_{EM}=&\tfrac12 \left[G_{Ep}\left(\tfrac14Q^2\right)G_{Mn}\left(\tfrac14Q^2\right) \right.\\
 &\left.- G_{En}\left(\tfrac14Q^2\right)G_{Mp}\left(\tfrac14Q^2\right)\right],\\
 G_{MM}=&G_{Mp}\left(\tfrac14Q^2\right)G_{Mn}\left(\tfrac14Q^2\right),\\
 G'_{MM}=&\tfrac23 G_{Mp}^2\left(\tfrac14Q^2\right)\\
 & + \tfrac13 G_{Mp}\left(\tfrac14Q^2\right)G_{Mn}\left(\tfrac14Q^2\right).
\end{split}
\ee

\mbox{}\\[-1.cm]

Finally we get
\be\label{S_00}
\begin{split}
 \mathfrak{M}_{00}=&\frac{3}{4\pi} \int_0^\infty d\rho
 \left\{
  G'_{EE}j_0\phi_1^2 
  \right.\\
 & \left. 
+ G_{EE}\left[j_0\left(\phi_2^2 + \phi_3^2 
\right)
 \right]
 \right\},
\end{split}
\ee

\mbox{}\\[-1.cm]
\be\label{S_0+}
\begin{split}
 \mathfrak{M}_{0+}=&\frac{\sqrt{\tau}}{\pi}  \int_0^\infty d\rho
  \left\{
  G'_{EM}\left(
               -\frac{j_0\phi_1\phi_2}{\sqrt2} 
               -\frac{j_2\phi_1\phi_3}{2} 
  \right)
  \right.\\
 & \left. 
+ G_{EM}\left[ \frac{ j_0\phi_2^2}{\sqrt2}  
              -\frac{j_2\phi_2\phi_3}{2} 
              -\frac{(j_0 - j_2) \phi_3^2}{2\sqrt2}
 \right]
 \right\},
\end{split}
\ee
\be\label{S_+-}
\begin{split}
 \mathfrak{M}_{+-}=&\frac{\tau}{\pi}  \int_0^\infty d\rho
 \left\{ -G'_{MM}\frac{3 j_0 \phi_1^2}4 \right.\\
 &\left. 
   + G_{MM}\left[ - \frac{j_0\phi_2^2}{4} 
                  + \frac{j_2 \phi_2\phi_3}{\sqrt2}  
\right. \right.\\
 &\left.\left. 
                  - \frac{(j_0-j_2)\phi_3^2}{4} 
 \right]
 \right\},
\end{split}
\ee
where we use the following notations
\be\label{Notations3}
\begin{split}
& \tau=\frac{-\Delta_n^2}{4m^2}\approx \frac{Q^2}{16m^2},\\
& j_\ell =j_\ell\left(\tfrac13 Q\rho\right),\\ 
& \phi_n=\phi_n(t,\rho).
\end{split}
\ee
%

\mbox{}\\[1.5cm]


\begin{thebibliography}{99}
\bibitem{Jlab.Jones}
M.K.~Jones et al., Phys. Rev. Lett., {\bf 84}, 1398 (2000).
\bibitem{Jlab.Gayou} 
O.~Gayou et al., Phys. Rev. Lett., {\bf 88}, 092301 (2002).
\bibitem{Punjabi}
V.~Punjabi et al., Phys. Rev. C {\bf 71}, 055202 (2005); ibid. {\bf 71}, 069902 (E) (2005).
\bibitem{Guichon}
P.A.M.~Guichon and M.~Vanderhaeghen, Phys. Rev. Lett., {\bf 91}, 142303 (2003).
\bibitem{BlundenPRL}
P.G.~Blunden, W.~Melnitchouk, and J.A.~Tjon, Phys. Rev. Lett., {\bf 91}, 142304 (2003).
\bibitem{Chen}
Y.C.~Chen, A.~Afanasev, S.J.~Brodsky, C.E.~Carlson, and M.~Vanderhaeghen, Phys. Rev. Lett., {\bf 93}, 122301 (2004).
\bibitem{BlundenPRC}
P.G.~Blunden, W.~Melnitchouk, and J.A.~Tjon, Phys. Rev. C {\bf 72}, 034612 (2005).
\bibitem{Kondratyuk}
S.~Kondratyuk, P.G.~Blunden, W.~Melnichouk, and J.A.~Tjon, Phys. Rev. Lett., {\bf 95}, 172503 (2005).
\bibitem{BK_PRC_2006}
D.~Borisyuk and A.~Kobushkin, Phys. Rev. C {\bf 74}, 065203 (2006).
\bibitem{BK_PRC_2007}
D.~Borisyuk and A.~Kobushkin, Phys. Rev. C {\bf 75}, 038202 (2007).
\bibitem{BK_PRC_2008}
D.~Borisyuk and A.~Kobushkin, Phys. Rev. C {\bf 78}, 025208 (2008).
\bibitem{BK_PRD_2009}
D.~Borisyuk and A.~Kobushkin, Phys. Rev. D {\bf 79}, 034001 (2009).
\bibitem{Kivel1}
N.~Kivel and M.~Vanderhaeghen, Phys. Rev. Lett., {\bf 103}, 092004 (2009) [arXiv:0905.0282].
\bibitem{Kivel2}
N.~Kivel and M.~Vanderhaeghen, JHEP, {\bf 1304}, 029 (2013) [arXiv:1212.0683].
\bibitem{Arrington}
J.~Arrington, Phys. Rev. C {\bf 71}, 015202 (2005).
\bibitem{Tvaskis}
V.~Tvaskis et al., Phys. Rev. C {\bf 73}, 025206 (2006).
\bibitem{Chen_Kao_Yang}
Y.-C.~Chen, C.-W.~Kao, and S.-N.~Yang, Phys. Lett. B {\bf 652}, 269 (2007) [arXiv:nucl-th/0703017].
\bibitem{BorisyukKob_phen}
D.~Borisyuk and A.~Kobushkin, Phys. Rev. C {\bf 76}, 022201 (R) (2007).
\bibitem{BK_PRD_2011}
D.~Borisyuk and A.~Kobushkin, Phys. Rev. D {\bf 83}, 057501 (2011).
\bibitem{Qattan}
I.A.~Qattan, A.~Alsaad, J.~Arrington, Phys. Rev. C {\bf 84}, 054317 (2011) [arXiv:1109.1441].
\bibitem{Dong_Wang}
Y.-B.~Dong and S.D.~Wang, Phys. Lett. B {\bf 684}, 123 (2010).
\bibitem{BlundenPRC_pion}
P.G.~Blunden, W.~Melnitchouk, and J.A.~Tjon, Phys. Rev. C {\bf 81}, 018202 (2010).
\bibitem{BK_PRC_2011_pion}
D.~Borisyuk and A.~Kobushkin, Phys. Rev. C {\bf 83}, 025203 (2011).
\bibitem{Dong_Chen}
Y.B.~Dong and D.Y.~Chen, Phys. Lett. B {\bf 675}, 426 (2009).
\bibitem{KK-ED}
A.P.~Kobushkin, Ya.D.~Krivenko-Emetov, and S.~Dubni\v cka, Phys. Rev. C {\bf 81}, 054001 (2010).
\bibitem{KK-EDD}
A.P.~Kobushkin, Ya.D.~Krivenko-Emetov, S.~Dubni\v cka, and A.-Z.~Dubni\v ckova, Phys. Rev. C {\bf 84}, 054007 (2011).
\bibitem{Boitsov}
V.N.~Boitsov, L.A.~Kondratyuk, and V.B.~Kopeliovich, Yad. Fiz. {\bf 16}, 515 (1972).
\bibitem{Baru}
V.~Baru, J.~Haidenbauer, C.~Hanhart, and J.A.~Niskanen, Eur. Phys. J. A {\bf 16}, 437 (2003).
\bibitem{Paris}
M.~Lacombe et al., Phys. Rev. C {\bf 21}, 861 (1980).
\bibitem{cd-Bonn}
R.~Machleidt, Phys. Rev. C {\bf 63}, 024001 (2001).
\bibitem{KobStrok}
A.P.~Kobushkin and E.A.~Strokovsky, Phys. Rev. C {\bf 87}, 024002 (2013) [arXiv:1204.0425].
\bibitem{TPEcalc}
D.L.~Borisyuk and A.P.~Kobushkin, arXiv:1209.2746v1 (2012).
\bibitem{PPV_2007}
C.F.~Perdrisat, V.~Punjabi, and M.~Vanderhaeghen, Prog. Part. Nucl. Phys. {\bf 59}, 694 (2007) [arXiv:hep-ph/0612014].
\bibitem{Galster}
S.~Galster et al., Nucl. Phys. B {\bf 32}, 221 (1971).
\bibitem{Amroun}
A.~Amroun et al., Nucl. Phys. A {\bf 579}, 596 (1994).
\end{thebibliography}
\end{document}